\documentclass[submission,copyright,creativecommons]{eptcs}
\providecommand{\event}{The 4th International Workshop on Symbolic and Numerical Methods\\ for Reachability Analysis, affiliated with ETAPS 2018} 
\usepackage{underscore}           
\usepackage{float}
\usepackage{makecell}
\usepackage{chngpage}
\usepackage{algorithm}
\usepackage[noend]{algpseudocode}
\usepackage[english]{babel}
\usepackage[utf8]{inputenc}
\usepackage{amsmath}
\usepackage{graphicx}
\usepackage[colorinlistoftodos]{todonotes}
\usepackage{amsfonts}
\usepackage{amssymb} 
\usepackage{lipsum}
\usepackage{blindtext}
\usepackage[utf8]{inputenc}
\usepackage{setspace}
\usepackage{amssymb}
\usepackage{graphicx}
\usepackage{multirow}
\usepackage{verbatim}
\usepackage{dirtytalk}
\usepackage{pdflscape}
\usepackage{afterpage}
\usepackage{rotating}
\usepackage{capt-of}
\usepackage{ctable} 
\def  \eg       {{\em e.g.}}
\def  \ie       {{\em i.e.}}

\title{Full version: An evaluation \\of estimation  techniques for  probabilistic reachability}

\author{Mariia Vasileva \qquad\qquad Paolo Zuliani
\institute{School of Computing, Newcastle University, UK}
\email{vasileva.m2@newcastle.ac.uk \quad\qquad paolo.zuliani@newcastle.ac.uk}
}

\def\titlerunning{An evaluation of estimation techniques for  probabilistic reachability}
\def\authorrunning{M. Vasileva \& P. Zuliani}
\begin{document}
\maketitle

\begin{abstract}
We evaluate numerically-precise Monte Carlo (MC), Quasi-Monte Carlo (QMC) and Randomised Quasi-Monte Carlo (RQMC) methods for computing probabilistic reachability in hybrid systems with random parameters. Computing reachability probability amounts to computing (multidimensional) integrals. In particular, we pay attention to QMC methods due to their theoretical benefits in convergence speed with respect to the MC method. The Koksma-Hlawka inequality is a standard result that bounds the approximation error of an integral by QMC techniques. However, it is not useful in practice because it depends on the variation of the integrand function, which is in general difficult to compute. The question arises whether it is possible to apply statistical or empirical methods for estimating the approximation error. In this paper we compare a number of interval estimation techniques based on the Central Limit Theorem (CLT), and we also introduce a new approach based on the CLT for computing confidence intervals for probability near the borders of the [0,1] interval.
Based on our analysis, we provide justification for the use of the developed approach and suggest usage guidelines for probability estimation techniques. 

\end{abstract}

\section{Introduction} \label{sec:Intro}

Reachability is one of the fundamental problems in verification and model checking. Given a system model and a set of \say{goal} states (indicating (un)wanted behaviour), does the system eventually reach these states? The generalisation of this problem for stochastic systems is called probabilistic reachability, and it amounts to compute the probability that the system reaches a goal state.

Checking reachability in hybrid systems is an undecidable problem for all but the simplest systems (timed automata) \cite{2}. Formal verification of hybrid systems can include verifying satisfiability of formulas involving real variables, which is known to be an undecidable problem when, \textit{e.g.}, trigonometric functions are involved \cite{22}. In order to combat the undecidability of general sentences over the reals, Gao, Avigad and Clarke defined the notion of $\delta$-complete decision procedure \cite{12}. 
This approach has been extended to a bounded probabilistic reachability method with statistically valid enclosures for the probability that a hybrid system can reach a goal state within a given time bound and number of steps \cite{20}.  In particular, we consider the $k$-step reachability probability for parametric hybrid systems with random parameters. 
The ProbReach tool \cite{19} implements such a method and computes under- and over- approximation of the reachability probability, which amounts to computing multi-dimensional integrals. 
There are three possible ways to compute such integrals - formal,  Monte-Carlo (MC) and Quasi-Monte Carlo (QMC). The number of system evolutions to explore in order to accurately compute integrals grows exponentially with respect to the number of dimensions \cite{22}. This motivates a combination of MC and QMC methods and numerical decision procedures in order to define efficient, numerically accurate estimation techniques. 

It is well-known that MC methods are based on the Law of Large Numbers and random sampling. Instead, QMC methods are based on {\em deterministic} sampling from so-called quasi-random sequences. The error estimation of the QMC method can be computed theoretically by the Koksma-Hlawka inequality. Unfortunately, in practice its use is connected with a number of calculation difficulties \cite{15}. The terms of quasi-random sequences are statistically dependent, so the Central Limit Theorem (CLT) can not be used for estimating the integration error. At the same time we can successfully use the CLT for estimating the error of {\em Randomised} Quasi-Monte Carlo (RQMC) methods.  

The aim of this paper is to compare different interval estimation techniques, in particular in the extreme cases of probability close to 0 or 1. A problem of many confidence interval (CI) techniques is that the actual coverage probability of the  interval near the boundaries (0 and 1) can be poor \cite{18,6}.

The paper is structured as follows: In Section 2, we briefly introduce probabilistic reachability for stochastic parametric hybrid systems. In Section 3, we present integral estimation methods including MC, QMC and RQMC. Additionally, we consider a recent approach to QMC variance calculation for statistical error estimation. In Section 4, we examine CI error estimation with approaches based on the standard CLT interval and on the $Beta$ function. In Section 5, we empirically compare those CI estimation techniques on five benchmarks, and derive usage guidelines. In Section 6, we provide conclusions and suggest future work in the area.  

\section{Probabilistic Reachability } \label{sec:ProbReach}

Hybrid systems provide a framework for modelling real-world systems that combine continuous and discrete dynamics \cite{2}. In particular parametric hybrid systems (PHSs) \cite{20} represent continuous and discrete dynamic behavior dependent on initial parameters, which remain unchanged during the system evolution. Such systems can both flow, described by a differential equation and jump, described by difference equations or control graphs.  

In order to combat the undecidability of reasoning over real sentences Gao {\em et al.}~\cite{12} defined $\delta$-complete decision procedures, which correctly decide whether a slightly relaxed sentence is satisfiable or not. 
Let $\delta \in \mathbb{Q^+} \cup \{0\}$ be a constant and $\phi$ a bounded $\Sigma_1$-sentence in the standard form: 
$\phi = \exists^{I_{1}} x_{1}, ..., \exists^{I_{n}} x_{n}: \bigwedge^{m}_{i = 1}(\bigvee^{k_i}_{j = 1} f_{ij}(x_{1}, ..., x_{n}) = 0) $,
where the $f_{ij}(x_{1}, ..., x_{n})$ are compositions of Type 2-computable functions (these are essentially \say{numerically computable} real functions, including transcendental functions and solutions of differential equations). 
The $\delta$-weakening of $\phi$ is the formula: 
$\phi^\delta = \exists^{I_{1}} x_{1}, ..., \exists^{I_{n}} x_{n}: \bigwedge^{m}_{i = 1}(\bigvee^{k_i}_{j = 1} |f_{ij}(x_1, ..., x_n)| \le \delta)$.
The bounded $\delta$-SMT problem asks for the following: given a sentence $\phi$ of the above form and $\delta \in \mathbb{Q^+}$, correctly decide one of the following:
\begin{itemize}
\item \textbf{unsat}: $\phi$ is false,
\item \textbf{$\delta$-true}: $\phi^\delta$ is true.
\end{itemize}
If the two cases overlap either decision can be returned. Standard bounded reachability questions over PHSs can be coded as $\Sigma_1$ sentences and \say{$\delta$-decided} by $\delta$-complete decision procedures \cite{14}.


In this paper we are concerned with {\em stochastic} PHSs, which introduce random parameters to an otherwise deterministic PHS. Bounded $k$-step reachability in PHSs aims at finding the {\bf probability} that for the given initial conditions, the system reaches a goal state in $k$ discrete transitions within a given finite time.
It can be shown that this probability can be computed  as an integral of the form $\int_{G} d\mathbb{P}$, where $G$ denotes the set of all random parameter values for which the system reaches a goal state in $k$ steps, and $\mathbb{P}$ is the probability measure associated with the random parameters \cite{20}.

\section{Integral Estimation Methods}  \label{sec:IntEst}

\subsection{Monte Carlo Method}  \label{sec:IntEst-MC}

Consider the integral $I=\int_{a}^{b} f(y) dy$, and a random variable $U$ on $[a,b]$. The expectation of $f(U)$ is:
$$
\mathbb{E}[f(U)]=\int_{a}^{b}  f(y) \varphi (y) dy
$$
where $\varphi$ is the density of $U$. If $U$ is uniformly distributed on $[a,b]$, then the integral becomes:
$$
I=\int_{a}^{b}  f(y) dy = (b-a) \mathbb{E}[f(U)].
$$
If we take $N$ points, uniformly distributed on $[a,b]$, and compute the sample mean $\frac{1}{N}\sum_{i=1}^{N}f(u_i)$, we obtain the MC integral estimation: 
\begin{equation}\label{eq:MC}
\int_{a}^{b}  f(y) dy \approx (b-a) \frac{1}{N} \sum_{i=1}^{N}f(u_i)
\end{equation}
According to the Strong Law of Large Numbers, this approximation  is convergent (for $N \to \infty$) to $I$ with probability one. 
The variance of the MC estimator (\ref{eq:MC}) is:
\begin{equation}\label{eq:MC-Var}
Var(MC)=\int_a^b...\int_a^b\bigg( \frac{1}{N}\sum_{i=1}^Nf(u_i)-I \bigg)^2 du_1...du_N=\frac{\sigma^2_f}{N}
\end{equation}
The MC integration error mean is  $\frac{\sigma_f}{\sqrt{N}}$, where $\sigma_f^2$ is the integrand variance, which is assumed to exist. In practice, the integrand variance is often unknown. That is why the next estimation is instead used:
\begin{equation*}
\widehat{\sigma}^2_f=\frac{1}{N-1}\sum_{i=1}^N\bigg(f(u_i)-\frac{1}{N}\sum_{i=1}^Nf(u_i) \bigg)^2 
\end{equation*}
This estimator possesses the unbiasedness property: $ \mathbb{E}[\widehat{\sigma}^2_f]=\sigma^2_f$. 

\subsection{Quasi-Monte Carlo Method} \label{sec:IntEst-QMC}

\begin{figure}
 \begin{center}
\includegraphics[height=5.6cm]{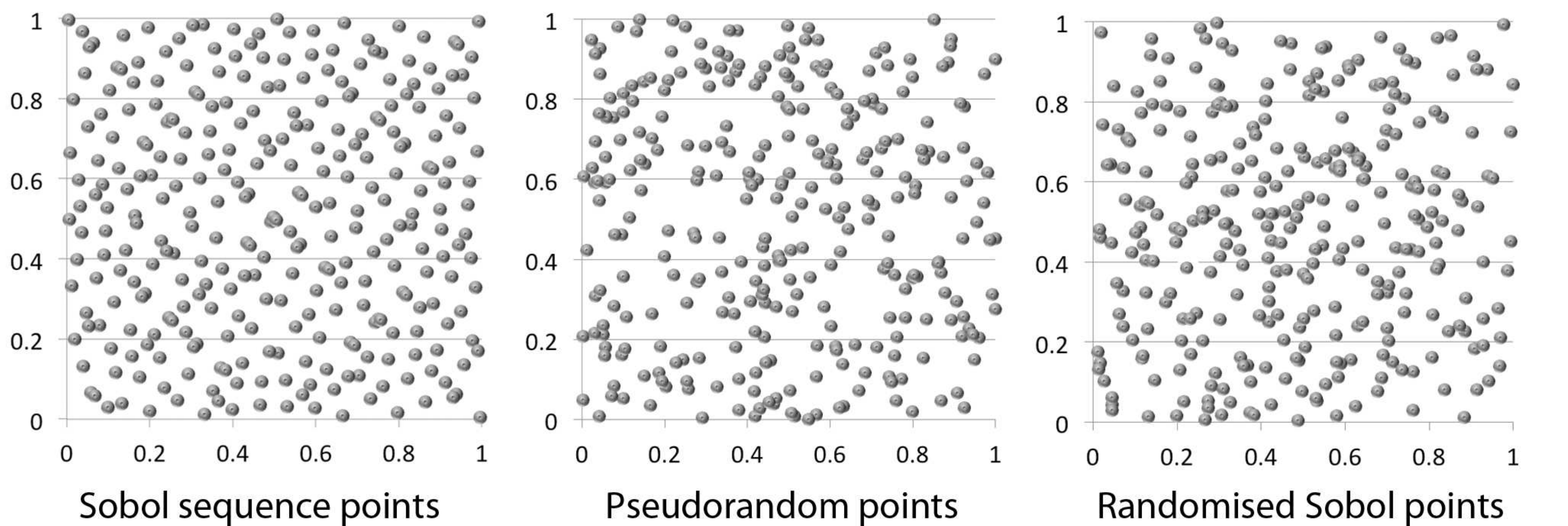}
 \end{center}
\caption{Sobol sequence, uniform pseudorandom and randomised Sobol sequence points (obtained by  transformation $\Gamma= (\mathfrak{X}+\epsilon)mod1$, where $\epsilon$ is a random sample from MC sequence and $\mathfrak{X}$ is low-discrepancy sample from Sobol sequence) distribution in the 2-dimensional unit space. The comparison is based on the first 300 points of sequences. } \label{fig:Sequence}
\end{figure}

QMC methods can be regarded as a deterministic counterpart to classical MC methods.  Unlike MC integration, which uses estimates (\ref{eq:MC}) with randomly selected points, QMC methods select the points $u_i$ {\em deterministically}. 
 Specifically, QMC techniques produce deterministic sequences of points that provide the best-possible spread over the integration domain. These deterministic sequences are often referred to as low-discrepancy sequences. The Sobol sequence \cite{21} is a well-known example of low-discrepancy sequence. In Figure \ref{fig:Sequence}, we present a simple example of the comparison between Sobol and pseudorandom points distribution. 
An effective way to use the QMC method is by performing a change of variables to reduce integration to the $[0,1]$ domain. When we need to integrate over a large domain $[a,b]$, that avoids multiplying the QMC estimate by a large factor $(b-a)$ as required by (\ref{eq:MC}). 

A QMC advantage with respect to MC is that its error is $O\left(\frac{1}{N}\right)$, while the MC error is  $O\left(\frac{1}{\sqrt{N}}\right)$, where $N$ is the sample size.
The Koksma-Hlawka inequality bounds the error of QMC estimates, but in practical applications it is very hard to estimate \cite{15}, thereby hampering the use of QMC methods. 
As such, other methods for estimating the QMC error need to be developed.

\paragraph{Qint Cubature Method.}  \label{sec:IntEst-Qint}
Ermakov and Antonov \cite{5} have recently introduced a new method for QMC variance estimation. 
To construct an estimate of the integral $I$ 
they use the set of random quadrature formulas, which were introduced by the Ermakov-Granovsky theorem \cite{11}. This theorem allows us to construct $N$-point formulas with two important properties: the unbiasedness property for integral $I$ and the accuracy property for the considered Haar system.
The nodes of the formula are random variables with distribution density:
\[ \phi (u_1, u_2,...,u_N)=
  \begin{cases}
    \frac{N^N}{N!}       & \quad \text{if }  \text{ $(u_1, u_2,...,u_N) \in Lat(i_1,i_2,...,i_N)$ }\\
    0       & \quad  \text{ otherwise}\\
  \end{cases}
\]
where $Lat(i_1,i_2,...,i_N)$ is a Latin set that relates to the permutation $(i_1,i_2,...,i_N)$ and can be defined by the next condition:
$$
(u_1, u_2,...,u_N) \in Lat(i_1,i_2,...,i_N)\Leftrightarrow \forall j \in \{1,2,...,N\}  u_j \in \mathcal{U}_{i_j}
$$
where $\mathcal{U}_{i_j}$ is a set of permuted orthonormal Haar functions \cite{5}.

The variance of the constructed cubature formula $Cub[f]=\frac{1}{N}\sum_{i=1}^N f(u_i)$
can be calculated as:
$$
\mathbb{D} Cub[f] =\int_\mathcal{U^N} Cub[f]^2d \phi- \bigg( \int_\mathcal{U^N} Cub[f] d \phi\bigg)^2=
$$
$$
=\mathbb{D} MC[f]+\frac{1}{N}(a_1+a_2+...+a_N)^2-a_1^2-a_2^2-...-a_N^2= \mathbb{D}_{MC}[f]-\frac{1}{N}\sum_{i<j}(a_i-a_j)^2,
$$
where 
$\mathbb{D} MC$ is the variance of MC method (\ref{eq:MC-Var}) and $a_i=\int_{\mathcal{U}_i} f(u)\mu (du)$  for $i=1,2,...,N$.

In other words, we can redefine the integral estimation variance as: 
\begin{equation}\label{eq:Qint-Var}
Var(QMC)=Var(MC)-\frac{1}{N}\sum_{i<j} (a_i-a_j)^2\ .
\end{equation}

\subsection{Randomised Quasi-Monte Carlo} \label{sec:IntEst-RQMC}

As discussed earlier, the practical application of QMC is limited by the difficulty of computing an estimate of the integration error. However, allowing randomization into the deterministic QMC procedure enables constructing CIs.
A {\em Randomised} QMC (RQMC) procedure can be described as follows. Suppose that  $\mathfrak{X}= \{x_1,...,x_n\}$ is a deterministic low-discrepancy set. By  means of a transformation $\tilde{\mathfrak{X}}=\Gamma (\mathfrak{X},\epsilon)$ a finite set $\tilde{\mathfrak{X}}$ is generated by the random variable $\epsilon$ and has the same quasi-random properties as set $\mathfrak{X}$ (see Figure 1).
For a randomised set $\tilde{\mathfrak{X}}_i$ we construct a RQMC estimate similar to (\ref{eq:MC}): 
\begin{equation}\label{eq:RQMC}
RQMC_{j,n}=\frac{1}{n}\sum_{i=1}^n f(\tilde{\mathfrak{X}}_{i,j})
\end{equation}
for $0<j \leqslant r$, where $r$ is the total number of different pseudo-random sequences. Then, we take their average for overall RQMC estimation (\ref{eq:RQMC}): 
\begin{equation}\label{eq:RQMC-av}
RQMC_{n}=\frac{1}{r}\sum_{j=1}^r RQMC_{j,n}.
\end{equation}
If we choose the $\Gamma$ transformation in such a way that each of the estimates $RQMC_{j,n}$ has the unbiasedness property, \textit{i.e.}, $\forall j$  $\mathbb{E}$ $[RQMC_{j,n}]=I$, (\textit{e.g.} $\Gamma= (\mathfrak{X}+\epsilon) $ mod $1$), then the estimator (\ref{eq:RQMC-av}) will also be unbiased,  \textit{i.e.}, $\mathbb{E} [RQMC_{n}]=I$.
By independence of the samples used in (\ref{eq:RQMC}) and (\ref{eq:RQMC-av}), we have that for all $0<j \leqslant r$:
\begin{equation*}
Var(RQMC_{n})=\frac{{Var(RQMC_{j,n})}}{r}.
\end{equation*}
Thus, we have the following variance estimation:
\begin{equation*}
\widehat{Var}(RQMC_{n})=\frac{1}{r(r-1)} \sum_{j=1}^r \Big( RQMC_{j,n}-RQMC_{n}\Big)^2\ .
\end{equation*}

\subsection{Validated MC and QMC} \label{sec:IntEst-Validation}

Let us consider a hybrid system $H$ with random parameters only. For any parameter value $p$ from the initial parameters distribution we introduce the Bernoulli random variable $X$, which takes 1 if system $H$ reaches the goal in $k$ steps for $p$ and 0 otherwise. Since in general we can not sample $X$ because of undecidability, we instead consider two Bernoulli random variables: $ X_{sat}$, which takes 1 if we can {\em correctly decide} that system $H$ reaches the goal in $k$ steps for $p$ and 0 otherwise; $ X_{usat}$, which takes 0 if we can {\em correctly} decide that system $H$ does not reach the goal and 1 otherwise \cite{20}. 
Therefore:
\[
X_{sat} \leqslant X \leqslant X_{usat}
\]
and thus:
\[
\mathbb{E}[X_{sat}]\leqslant \mathbb{E}[X] \leqslant \mathbb{E}[X_{usat}] \ .
\]
By the definition of expectation, and denoting $P_R$ as the domain of the random parameters of $H$,
we get:
\begin{equation}\label{ineq:Val-Exp}
\int_{P_R} X_{sat}(p) dp \leqslant \int_{P_R} X(p) dp \leqslant \int_{P_R} X_{usat}(p) dp\ .
\end{equation}
We take the sample approximation of (\ref{ineq:Val-Exp}) and obtain
\[
\frac{1}{N} \sum_{i=1}^N X_{sat} (p_i)   \leqslant \frac{1}{N} \sum_{i=1}^N X(p_i) \leqslant \frac{1}{N} \sum_{i=1}^N X_{usat} (p_i) 
\]
where the $p_i$'s can be sampled by using low-discrepancy sequences for QMC methods or pseudo-random sequences
for MC methods.

\section{Confidence Interval Estimation and Error Analysis} \label{sec:CI}

In the following we shall use the notation below: \\
\textcolor{white}{_} $\bullet$ $\tilde{X}=\frac{1}{n}\sum_{i=1}^n x_i$ - the sample mean. \\
\textcolor{white}{_} $\bullet$ $C_a=Quant(1-\frac{a}{2})$ - the inverse cumulative distribution function of a normal random variable with \textcolor{white}{_l}mean 0 and standard deviation 1; parameter $a$ defines the confidence level at $1-a$. \\
\textcolor{white}{_} $\bullet$ $\hat{p}=n_s/n$ - the binomially-distributed proportion, where: $n_s$ - the number of \say{successes} and $n_f$ - the \textcolor{white}{_} number of \say{failures} in Bernoulli trial process; $n$ - the total number of Bernoulli \say{trials}.\\
\textcolor{white}{_} $\bullet$ $\hat{q} = 1-\hat{p}$.

\subsection{Intervals Based on the Standard CLT Interval}\label{sec:CLT-CI}

\paragraph{Modified Central Limit Theorem interval.}
First, we consider the case when the sample $x_i$ is extracted from the normal distribution $ N (\mu, \sigma^2)$ with unknown parameter $\mu$ and known $\sigma^2$, where $\mu$ is the mean or expectation of the distribution and $\sigma^2$ is the variance. Here, $\mu$  can be approximated by the sample mean: $\mu \approx \tilde{X}$.
To clarify this approximation, we construct a CI covering the parameter $\mu$ with a given confidence probability:
\begin{equation}\label{CI:CLT}
CI_{CLT}=\left(\tilde{X}-C_a \frac{\sigma}{\sqrt{n}};\tilde{X}+C_a \frac{\sigma}{\sqrt{n}}\right)
\end{equation}
If the variance $\sigma^2$ is unknown, we can use the same CI by replacing $\sigma$ with the sample standard deviation
$s=\sqrt{\frac{1}{n-1}\sum_{i=1}^n(x_i-\tilde{X})^2}$. 
This method is widely used for estimating the distribution of the error regarding the binomially-distributed proportions. 
Many related works \cite{6,7,8} note that the $CI_{CLT}$ approximation can be poor when applied to Bernoulli trials with $\hat{p}$ close to 0 or 1.

In order to resolve this problem, we introduce a new method for variance estimation, which uses a sequential estimation of the sample standard deviation and calculates $CI_{CLT}$ (\ref{CI:CLT}) at every new sample. Our solution simply approximates the sample standard deviation with $\frac{1}{n^2}$ at the initial stages of the computation if $\hat{p}$ is equal to 0 (or 1) and propagates it through the computation until the necessary number of samples to construct the interval are obtained. We show the advantages of this approach in Section \ref{sec:Results}.

\paragraph{Wilson interval.}
It was introduced by Wilson in 1927 in his fundamental work \cite{9} and uses the inversion of the CLT interval. Additionally, it involves a modified center by quantile formula mean value. The interval has the following form:
\begin{equation}\label{CI:Wilson}
CI_{W}=\left(\frac{n_s+\frac{C_a^2}{2}}{n+C_a}-\frac{C_a \sqrt{n}}{n+C_a^2}\sqrt{\hat{p}\hat{q}+\frac{C_a^2}{4n}};\frac{n_s+\frac{C_a^2}{2}}{n+C_a}+\frac{C_a \sqrt{n}}{n+C_a^2}\sqrt{\hat{p}\hat{q}+\frac{C_a^2}{4n}}\right).
\end{equation}
This interval has some obvious advantages - it can not exceed the probability boundaries, and it can be easily calculated even if $\hat{p}$ is 0 or 1. At the same time, $CI_W$ has downward spikes when $\hat{p}$ is close to 0 and 1, because it is formed by an inverted CLT approximation. 

\paragraph{Agresti-Coull interval.}
This method was introduced by Agresti and Coull in 1998 \cite{1}. One of the most interesting features of this CI is that it makes a crucial assumption about $n_s$ and $n_f$. This interval formally adds two successes and two failures to the obtained values in case of 95\% confidence level and then uses the CLT method. The interval can be constructed as follows:
\begin{equation}\label{CI:AC}
CI_{AC}=\left(\tilde{X}- \frac{1}{n+C_a^2}(n_s+\frac{1}{2}C_a^2);\tilde{X}+\frac{1}{n+C_a^2}(n_s+\frac{1}{2}C_a^2)\right)
\end{equation}
Additionally, this interval can be modified by using the center of the Wilson interval (\ref{CI:Wilson}) in place of $\hat{p}$:
\begin{equation}\label{CI:AC_W}
CI_{AC_W}=\left(\frac{n_s+\frac{C_a^2}{2}}{n+C_a}-C_a\sqrt{\hat{p}\hat{q}(n+C_a^2)};(\frac{n_s+\frac{C_a^2}{2}}{n+C_a}-C_a\sqrt{\hat{p}\hat{q}(n+C_a^2)}\right)\ .
\end{equation}

\paragraph{Logit interval.}
The Logit interval is based on a transformation of the standard interval \cite{8}. It uses the empirical logit transformation:  $\lambda=ln(\frac{\hat{p}}{1-\hat{p}})=ln(\frac{n_s}{n-n_s})$. The variance of $\lambda$ is:
$ \widehat{Var}(\lambda)=\frac{n}{n_s(n-n_s)}$ and the Logit interval can be estimated as:
\begin{equation}\label{CI:Logit}
CI_{L}=\left(\frac{e^{\lambda_L}}{1+e^{\lambda_L}},\frac{e^{\lambda_U}}{1+e^{\lambda_U}}\right)
\end{equation}
where the lower bound transformation is $\lambda_L=\lambda-C_a\sqrt{\widehat{Var}(\lambda)}$ and the upper bound transformation is $\lambda_U=\lambda+C_a\sqrt{\widehat{Var}(\lambda)}$.

\paragraph{Anscombe interval.}
This interval was proposed by Anscombe in 1956 \cite{4} and is based on the Logit interval (\ref{CI:Logit}). The key difference is in $\lambda$ and $\widehat{Var}(\lambda)$ estimation, where $\lambda$ is defined as
$ \lambda=ln(\frac{n_s+\frac{1}{2}}{n-n_s+\frac{1}{2}}) $ and the variance is
$
\widehat{Var}(\lambda)=\frac{(n+1)(n+2)}{n(n_s+1)(n-n_s+1)}.
$
On this basis, the Anscombe interval $CI_{Anc}$ is estimated in the same way as the Logit interval (\ref{CI:Logit}).

\paragraph{Arcsine interval.}

It uses a variance-stabilising transformation of $\hat{p}$. In 1948, Anscombe introduced an improvement \cite{3} for achieving better variance stabilisation by replacing $\hat{p}$ to $p^\dagger=\frac{n_s+3/8}{n+3/4}$,
obtaining
\begin{equation}\label{CI:Arcsine}
CI_{Arc}=\left(\sin(\arcsin(\sqrt{p^\dagger})-\frac{C_a}{2\sqrt{n}})^2,\sin(\arcsin(\sqrt{p^\dagger})+\frac{C_a}{2\sqrt{n}})^2\right) \ .
\end{equation}

\subsection{Alternative Intervals Based on the Beta-Function} \label{sec:Beta-CI}


\paragraph{Bayesian interval.}
This method is based on the assumption that the (unknown) probability $p$ to estimate is a random quantity \cite{24}. The Bayesian interval is also called \say{credible}, because it computes the posterior distribution of the unknown quantity by using its prior distribution and Bayes theorem. The prior distribution can be constructed by means of the $Beta$ distribution, which is widely used for computing inferences on $p$.
If $p$ has a prior distribution $Beta(\alpha,\beta)$ then $p$ has posterior distribution  $Beta(n_s+\alpha,n-n_s+\beta)$. We can construct a Bayesian equal-tailed interval by the formula:
\begin{equation}\label{CI:Bayesian}
CI_B=\left(Beta^{-1}(\frac{a}{2},n_s+\alpha,n-n_s+\beta),Beta^{-1}(1-\frac{a}{2},n_s+\alpha,n-n_s+\beta)\right)
\end{equation}
where, $Beta^{-1}(a,\alpha,\beta)$ is the inverse of the cumulative distribution function of $Beta(\alpha,\beta)$.

\paragraph{Jeffreys interval.}

The Jeffreys interval is a Bayesian interval and uses the Jeffreys prior \cite{16}, which involves a non-informative prior given by the $Beta$ distribution with parameters $(\frac{1}{2}, \frac{1}{2})$. The probability density function of the $Beta$ distribution is
$
f(x;\alpha, \beta)=\frac{1}{B(\alpha, \beta)}x^{\alpha-1}(1-x)^{\beta-1},
$
where $0 \leq x \leq 1$, $\alpha, \beta>0$ and $B$ is beta function.
We can form the Jeffreys equal-tailed interval by (\ref{CI:Bayesian}) with parameters $(\frac{1}{2}, \frac{1}{2})$.


\paragraph{Clopper-Pearson interval.}
This method was introduced by Clopper and Pearson in 1934 \cite{7} and is based on the inversion of binomial test, rather than on approximations. 
The Clopper-Pearson interval is:
\begin{equation}\label{CI:CP}
CI_{CP}=\left(Beta^{-1}(\frac{a}{2},n_s,n-n_s+1),Beta^{-1}(1-\frac{a}{2},n_s+1,n-n_s)\right)\ .
\end{equation}
The $CI_{CP}$ interval states that the computed coverage probability is always above or equal to the $1-a$ confidence level. In practice, it can be achieved in cases when $n$ is large enough, while in general, the actual coverage can exceed $1-a$. We can conclude from equation (\ref{CI:CP}) that due to the absence of the $\alpha$ and $\beta$ parameters, the appropriate result can be achieved only by increasing number of \say{trials}. 

\section{Results}\label{sec:Results} \label{sec:Res}

We apply CI estimation methods, based on the standard interval with the RQMC technique and Bayesian CI estimation method with the MC technique. 
All our results are the average of 10 runs: in the RQMC case 10 sequences were obtained by changing the pseudo-random points $\epsilon$ of the equation $\Gamma= (\mathfrak{X}+\epsilon)mod1$, while the Sobol sequence points $\mathfrak{X}$ remain the same; in the MC case we used the same 10 pseudo-random points sequences, which were used for RQMC calculation.

\subsection{Border Probability Cases} \label{sec:Res-border}

The true probability values, 
which are shown in Section \ref{sec:Res-border} and Section \ref{sec:Res-MC/QMC} were obtained via pseudo-random number generation that produces boolean values according to a Bernoulli distribution.

\begin{figure} 
 \begin{center}
\includegraphics[height=23cm]{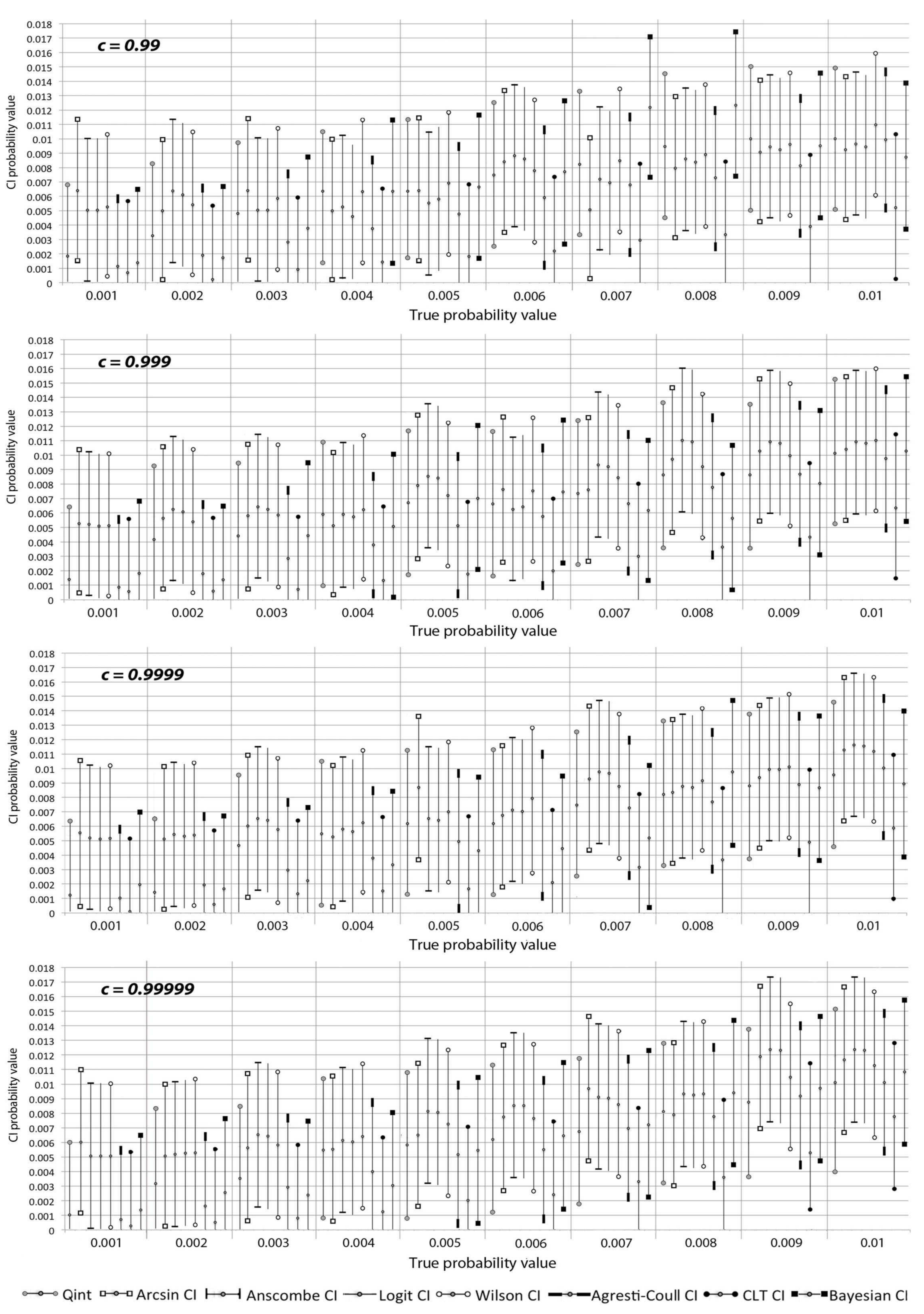}
 \end{center}
\caption{Comparison of confidence interval distribution for probability values near 0, interval size equal to $10^{-2}$ and  \textbf{\textit{c}} - confidence level.} \label{fig:Intervals}
\end{figure}

\paragraph{Intervals based on CLT and Bayesian interval.}
The comparison of the different CIs estimation techniques for extreme probability cases (near 0 bound) with accuracy $\epsilon=5 \times 10^{-3}$, which is presented in Figure \ref{fig:Intervals}, shows that all intervals except the Arcsin interval (\ref{CI:Arcsine}) (see plot $c=0.99$ of Figure \ref{fig:Intervals} for probability=0.001) contain the true probability value within their bounds. The Bayesian method tends to overestimate the true probability values according to their increase while $CI_{CLT}$ tends to underestimate them. Also, it is interesting to note that the most accurate center value is returned by the Agresti-Coull interval. The reason why $CI_{CLT}$ tends to include the true probability value near the upper bound of the interval is directly related to the number of samples. As it is shown in Figure \ref{fig:Intervals} for true probability values 0.007 - 0.01, the $CI_{CLT}$ center is moving up evenly to the true probability value with the increase of the confidence value. 
It echoes the number of samples growth for obtaining the necessary confidence level. For the other true probability values (0.001-0.006), although this trend retained, it can not be seen from the Figure, because of the small difference in the number of samples for all confidence levels, which causes the CI center to move wave-like.  

The results in Figure \ref{fig:Intervals} also demonstrate that the CIs based on the standard interval can have interval size smaller than its nominal value even for \say{large} sample sizes. It can be seen that every confidence level from 0.99 to 0.99999 displays further instances of the inadequacy of the CIs size.
Also, Figure \ref{fig:Intervals} shows that in spite of the \say{large} sample size (according to the confidence level), the size of the $CI_{B}$, $CI_{CLT}$ and $CI_{AC_w}$ intervals decreases significantly as $p$ moves toward 0.
Also, CIs based on the standard interval have interval size changes because of two reasons: absence of a posteriori estimate and skewness of the underlying binomial distribution. The erratic and unsatisfactory coverage properties of the standard interval have often been remarked on, but still do not seem to be widely appreciated among statisticians \cite{6}. The sequence sampling version still has the same disadvantages.

\begin{figure}[p] 
 \begin{center}
\includegraphics[height=12.6cm]{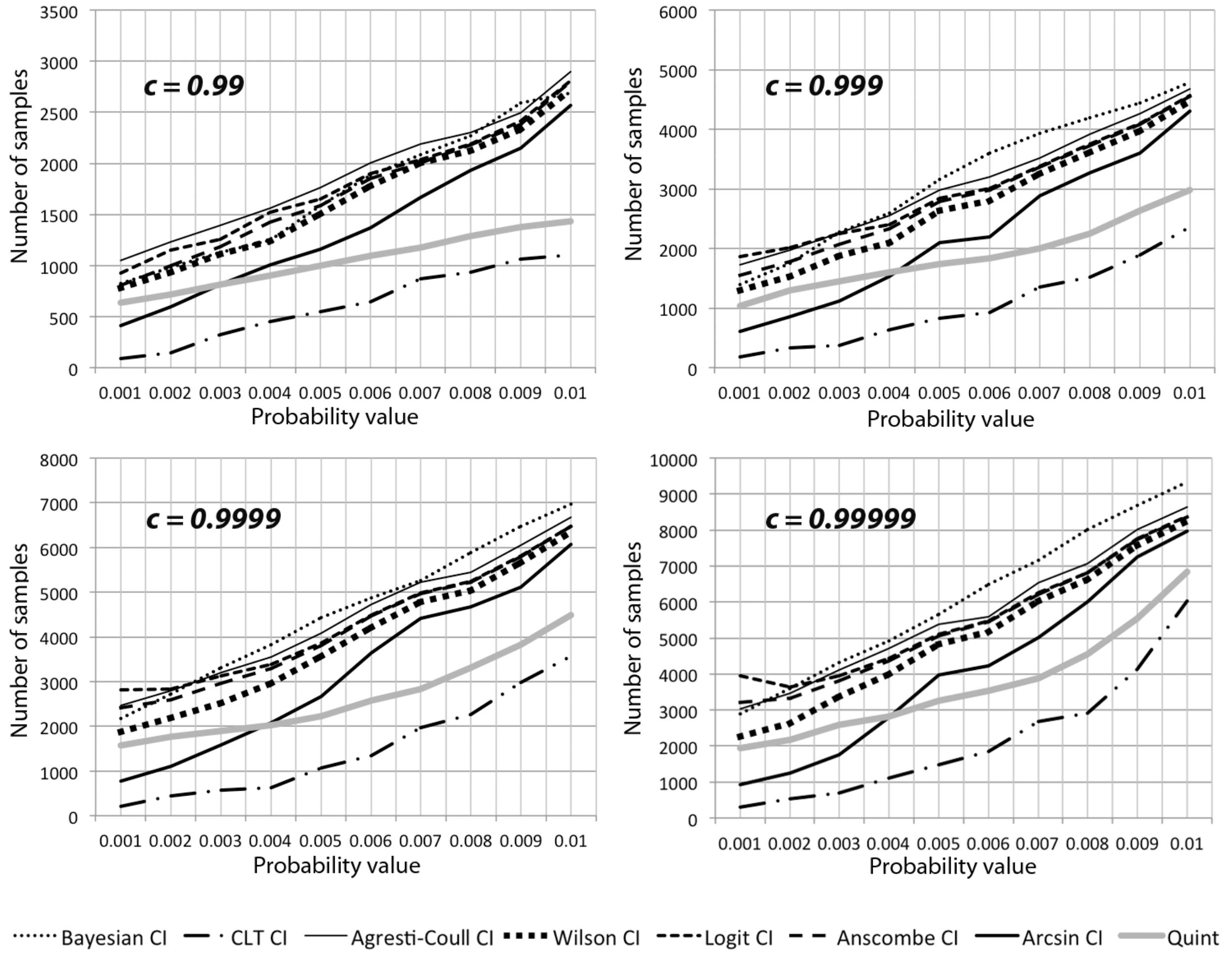}
 \end{center}
\caption{Comparison of sample size for probability values near 0, interval size equal to $10^{-2}$ and  \textbf{\textit{c}} - confidence level. } \label{fig:Points}
\end{figure}

In Figure \ref{fig:Points} we plot the number of samples that different CI estimation techniques used to return intervals with accuracy $\epsilon=5 \times 10^{-3}$ for different confidence levels. It can be clearly seen from the plots that with the increasing of the confidence all CIs based on the standard interval outperform the Bayesian CI. The plot with $c=0.99999$ in Figure \ref{fig:Points} displays that the best techniques in the number of samples from the best to the worst are: $CI_{CLT}$,  $Qint$, $CI_{Arc}$,  $CI_{W}$, $CI_{L}$, $CI_{Ans}$, $CI_{AC_W}$  and $CI_{B}$. The $CI_{L}$ and $CI_{Anc}$ techniques always show almost the same results near the bounds, because of the modification of the $CI_{L}$. Initially,  $CI_{L}$ is not able to deal with probability values near the bounds according to its $\lambda$ formula (see Section \ref{sec:CLT-CI}). It has been modified to use the Anscombe estimation formula in cases when  $\hat{p}=0$ or $\hat{p}=1$. It is also important to note that the difference in sample number between $CI_{CLT}$, $CI_{Arc}$ and $CI_{B}$ for extreme probability cases is significant. For example in the  plot with $c=0.9999$ of Figure \ref{fig:Points} the number of samples used to obtain interval for $p=0.005$ equals to 1,078 for $CI_{CLT}$, 2,662 for the $CI_{Arc}$ and 4,440 for $CI_{B}$.  

\begin{figure}[p] 
 \begin{center}
\includegraphics[height=5.2cm]{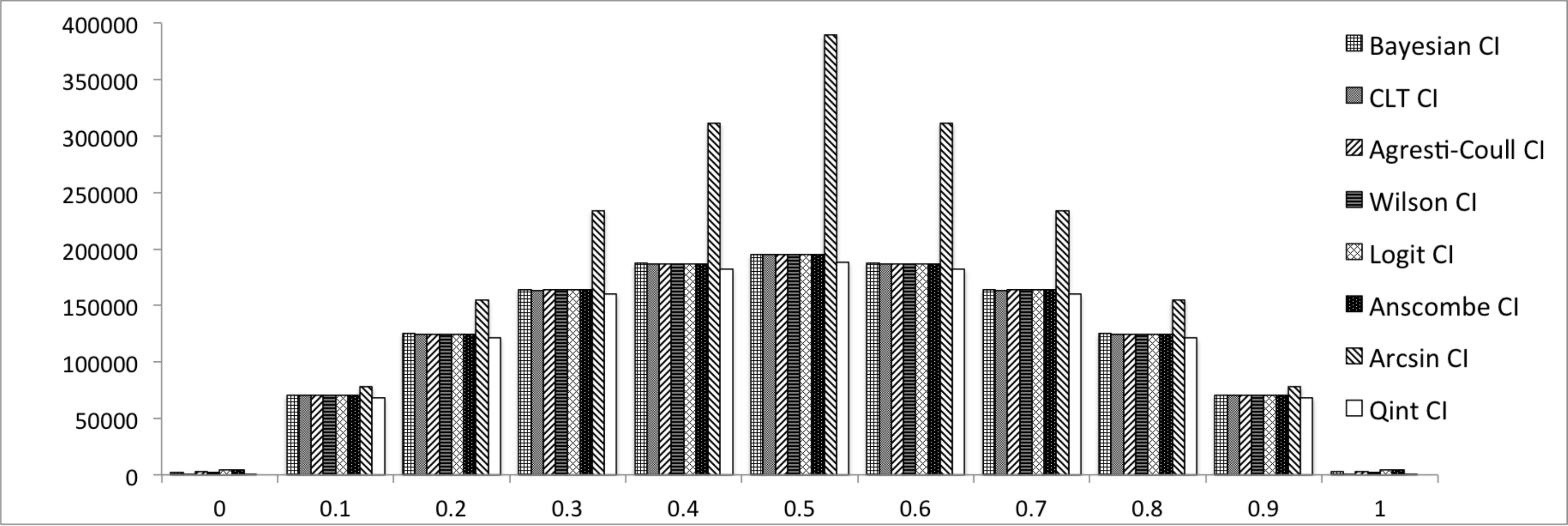}
 \end{center}
\caption{Comparison of sample size for probability values from 0 to 1, interval size equal to $10^{-2}$ and  confidence level equal to 0.99999. } \label{fig:Probability}
\end{figure}

This trend is not preserved with the increase of the probability value from 0 to 0.5 and with the decrease from 1 to 0.5, respectively. Figure \ref{fig:Probability} shows that the difference in sample number between all CIs (except $CI_{Arc}$) is almost undetectable. At the same time,  $CI_{Arc}$ shows very \say{bad} results in comparison with the others, as opposed to its results for probability values at the extremes.

Summarizing, for probability values near the bounds (0 or 1) the modified CLT method achieves better results in number of samples in comparison with the others (see Figure \ref{fig:Points}). For probability values away from the bounds, CLT, Wilson, Agresti–Coull, Logit and Anscombe methods are all very similar (see Figure \ref{fig:Probability}), and so for such probabilities we come to the conclusion that the CLT interval should be recommended, due to its simplest form. Meanwhile for smaller sample sizes, the $CI_{CLT}$ is strongly preferable to the others and so might be the
choice where sampling cost is paramount.

\paragraph{Qint method results.}

In Figure \ref{fig:Intervals} and Figure \ref{fig:Points} we also plotted the results of the recently developed Qint algorithm \cite{5}. In our research we used Qint with $n=k \times 2^s$, where $k=2$. These parameters were used to form $n$ points of the Sobol sequence $x_i$ with numbers $i \in I_{k,s}=\{1,2,...,k \times 2^s\}$.  
These parameters were chosen on the basis of the original study of the Qint method as the most universal and reliable. 
As it was described earlier, Qint uses a cubature randomization method and provides the integral estimation variance 
(\ref{eq:Qint-Var}). This formula is used to obtain a CI by calculating the standard interval (\ref{CI:CLT}) with our modification. 

In Figure \ref{fig:Intervals} we display the Qint intervals distribution for border probability values. We can see from the plots that the Qint CI always contains the true probability value. At the same time for all confidence levels from 0.99 to 0.99999 and for true probability values 0.006-0.01, Qint shows better centration than $CI_{B}$ and $CI_{CLT}$. The greatest differences between the Qint CLT center result and the true probability values are: 0.00245 for $c=0.99$ (p=0.004), 0.00191 for $c=0.999$ (p=0.004), 0.00168 for $c=0.9999$ (p=0.003), 0.00141 for $c=0.99999$ (p=0.004), while for example this difference for  $CI_{B}$ reaches 0.00518 for $c=0.99$ (p=0.007), 0.00235 for $c=0.999$ (p=0.008), 0.00181 for $c=0.9999$ (p=0.008), 0.00143 for $c=0.99999$ (p=0.008).

We can see in Figure \ref{fig:Points} that, as it was expected, Qint uses fewer samples than other CIs but $CI_{CLT}$.  
Our modification allows the Qint algorithm to return intervals even if $n_s=0$, which  significantly decreases the final sample size for 10 runs.
With the increase in $n$, which leads to further sampling and better reflects the behavior of the underlying random process, the effectiveness of the method decreases, and the benefit no longer seem so significant. 

The fact that with the chosen parameters Qint can not outperform our modified $CI_{CLT}$ leads us to the conclusion that our use of the standard deviation formula with $\frac{1}{n^2}$ lower bound is a rather effective and simple solution. However, the deep range of the possible parameters variation as well as novelty of the Qint algorithm lead us to believe that further research towards their comparison is needed.

\subsection{MC and QMC Error Comparison} \label{sec:Res-MC/QMC}

\begin{figure}[p]  
    \begin{adjustwidth}{-.6in}{-.6in}  
   \begin{minipage}{0.57\textwidth} 
     \centering
     \includegraphics[width=1\linewidth]{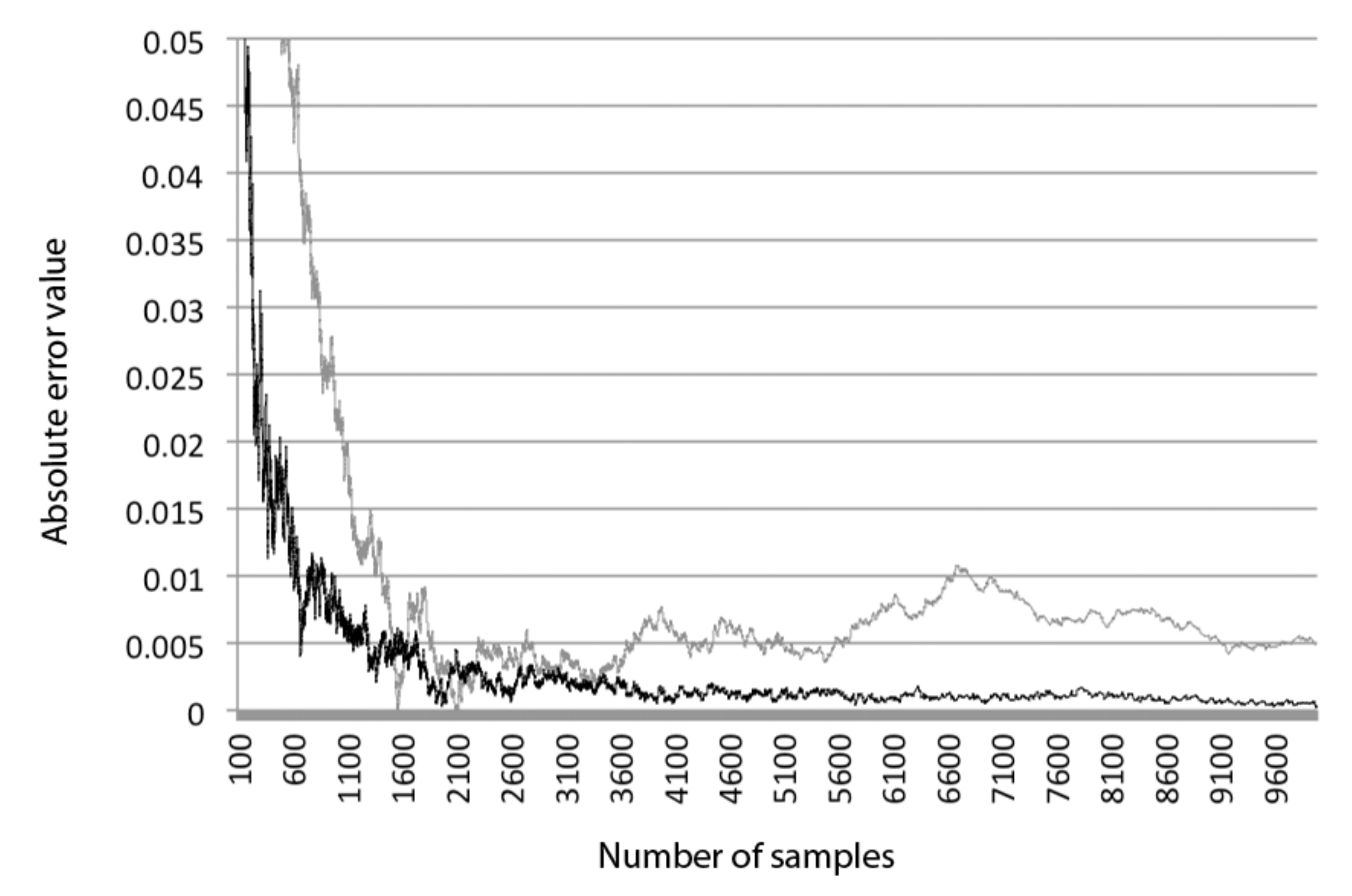}
     \caption{MC (grey line) and QMC (black line) absolute error with respect to the number of samples. Model: Collision advanced, type - max.}  \label{fig:ColMax}
   \end{minipage}\hfill
   \begin{minipage}{0.57\textwidth} 
     \centering
 	\includegraphics[width=1\linewidth]{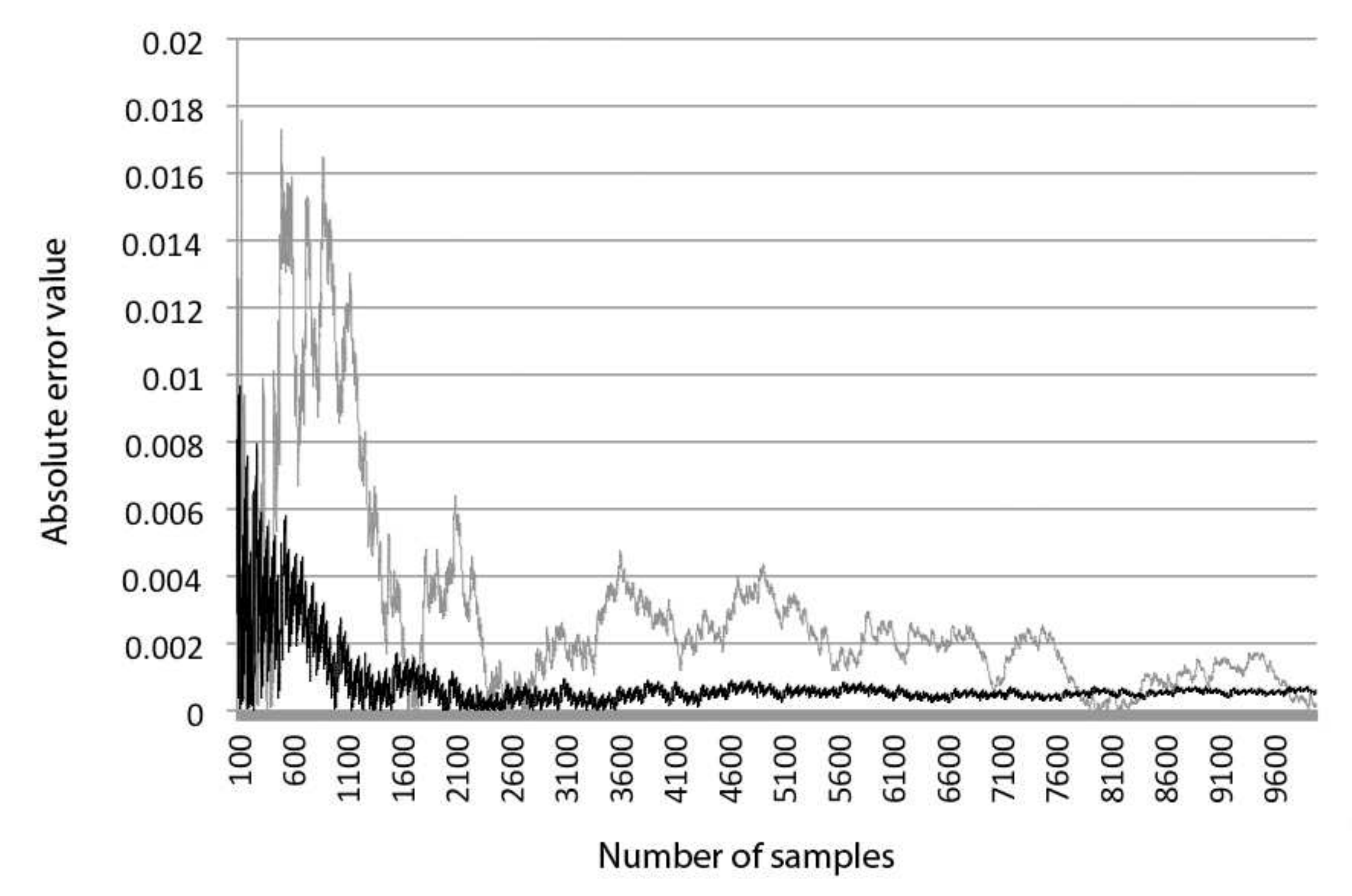}
    \newcommand{\tab}[1]{\hspace{.9\textwidth}}
     \caption{MC (grey line) and QMC (black line) absolute error with respect to the number of samples. Model: Bad, \tab ttype - max2.}  \label{fig:BadMax2}
   \end{minipage}
    \end{adjustwidth}
\end{figure}

Another key difference between the Bayesian CI and the CIs based on CLT is the use of MC and QMC techniques for interval calculation. As it was described in Section \ref{sec:IntEst-QMC}, the QMC advantage in the error size holds for all of the tested models (see Figures 5-16). In the cases where the true error rate could not be detected due to the probability value extremely close to 0 (\say{Bad} model type min and Collision (Basic) model type min), 
we have that the MC absolute error line equals the true probability value, because $n_s=0$ was obtained. 
The chaotic coverage properties of the MC method are far more persistent than they are appreciated. The chaotic behavior does not disappear even when $n$ is quite large and the true probability $p$ is not near the boundaries. For instance, in Figure \ref{fig:ColMax} it is visible that even when $n$ is quite large (\ie , tends to 10,000 samples) the actual absolute error value of the MC method reaches $5 \times 10^{-3}$. Hence we can conclude that CIs estimation techniques based on MC are misleading and defective in several respects and should not be trusted \cite{6}.

A notable phenomenon, which was noticed for the MC and QMC probability calculation is that the actual coverage probability contains non-negligible oscillations as both $p$ and $n$ vary. There exist some
“unlucky” pairs $(p, n)$ such that the corresponding absolute error is much greater than the results for smaller $n$. The phenomenon of oscillation is both in $n$, for fixed $p$, and in $p$, for fixed $n$. Furthermore, drastic
changes in coverage occur in nearby $p$ for fixed $n$ and in nearby $n$ for fixed $p$ \cite{6}. We can see it on the simple example in Figure \ref{fig:BadMax2}.


\begin{sidewaystable}[p] 
	\begin{adjustwidth}{-1.3in}{-.3in}  
     \begin{landscape}
        \begin{center}
        \scriptsize 
		\label{tab:simParameters}
		\begin{tabular}{|c|c|c|c|c|c|c|c|c|c|c|c|}
            
        \specialrule{.12em}{.05em}{.05em}    
        \multicolumn{11}{|c|}{\textbf{\textit{Confidence level c=0.99}}}\\ 
    	\specialrule{.12em}{.05em}{.05em}
			 \textbf{Model}  & \textbf{Type} & \textbf{\textit{P}} & \textbf{$CI_B$} & \textbf{$CI_{CLT}$} & \textbf{$CI_{AC_W}$}& \textbf{$CI_W$} & \textbf{$CI_L$}& \textbf{$CI_{Ans}$}& \textbf{$CI_{Arc}$} & \textit{Qint} \\
			\specialrule{.1em}{.05em}{.05em}
         \multirow{2}{*}{Good} & max&0.1& \makecell{$[$0.09671,  0.10671$]$}  & \makecell{$[$0.09564,  0.10564$]$} & \makecell{$[$0.09632,  0.10632$]$}  & \makecell{$[$0.09574,  0.10574$]$}  & \makecell{$[$0.09575,  0.10575$]$}  & \makecell{$[$0.09577,  0.10577$]$}  & \makecell{$[$0.09559,  0.10559$]$} & \makecell{$[$0.09147,  0.10147$]$}\\  
                               & min&0.1& \makecell{$[$0.09529,  0.10529$]$}  & \makecell{$[$0.0956,  0.1056$]$} & \makecell{$[$0.09666,  0.10666$]$}  & \makecell{$[$0.09679,  0.10679$]$}  & \makecell{$[$0.09678,  0.10678$]$}  & \makecell{$[$0.0968,  0.1068$]$}  & \makecell{$[$0.09639,  0.10639$]$} & \makecell{$[$0.09164,  0.10164$]$}\\ 
                               \hline
         \multirow{3}{*}{Bad} & max&0.95001& \makecell{$[$0.94416,  0.95416$]$}  & \makecell{$[$0.94495,  0.95493$]$} & \makecell{$[$0.94422,  0.95422$]$}  & \makecell{$[$0.94397,  0.95397$]$}  & \makecell{$[$0.94396,  0.95396$]$}  & \makecell{$[$0.94392,  0.95392$]$}  & \makecell{$[$0.94735,  0.95735$]$} & \makecell{$[$0.94459,  0.95459$]$} \\  
                              & max2&0.88747& \makecell{$[$0.8825,  0.8925$]$}  & \makecell{$[$0.88028,  0.89028$]$} & \makecell{$[$0.88031,  0.88031$]$}  & \makecell{$[$0.88019,  0.89019$]$}  & \makecell{$[$0.8803,  0.8902$]$}  & \makecell{$[$0.88019,  0.89019$]$}  & \makecell{$[$0.88325,  0.89325$]$} & \makecell{$[$0.88136,  0.89136$]$}\\                                    & min&$4 \times 10^{-7}$&
\makecell{$[$0,  0.00525$]$}  & \makecell{$[$0,  0.005$]$} & 
\makecell{$[$0,  0.00483$]$}  & \makecell{$[$0,  0.00955$]$}  & \makecell{$[$0.00005,  0.00959$]$}  & \makecell{$[$0.00005,  0.00959$]$}  & \makecell{$[$0.00131,  0.00959$]$}& [0,0.005] \\ 
                               \hline
         \multirow{2}{*}{\makecell{Deceleration}} & max& [0.08404,  0.08881] &
\makecell{$[$0.08471,  0.09471$]$}  & \makecell{$[$0.08802,  0.09802$]$} & \makecell{$[$0.08817,  0.09817$]$}  & \makecell{$[$0.08685,  0.09685$]$}  & \makecell{$[$0.08614,  0.09614$]$}  & \makecell{$[$0.0863,  0.0963$]$}  & \makecell{$[$0.08963,  0.09932$]$} & \makecell{$[$0.08852,  0.09852$]$} \\  
                & min&[0.04085,  0.04275]& 
\makecell{$[$0.03835,  0.04835$]$}  & \makecell{$[$0.03861,  0.04861$]$} & \makecell{$[$0.03854,  0.04854$]$}  & \makecell{$[$0.03884,  0.04884$]$}  & \makecell{$[$0.03886,  0.04886$]$}  & \makecell{$[$0.0389,  0.0489$]$}  & \makecell{$[$0.03873,  0.04873$]$}  & \makecell{$[$0.03337,  0.04337$]$}  \\    
                                  \hline
         \multirow{2}{*}{\makecell{Collision \\ (Basic)}} & max&[0.96567,  0.97254]& 
\makecell{$[$0.96371,  0.97381$]$}  & \makecell{$[$0.96873,  0.97873$]$} & \makecell{$[$0.9684,  0.9784$]$}  & \makecell{$[$0.96851,  0.97851$]$}  & \makecell{$[$0.96875,  0.97875$]$}  & \makecell{$[$0.96853,  0.97853$]$}  & \makecell{$[$0.96851,  0.97851$]$} & \makecell{$[$0.96301,  0.97301$]$}\\   
                   & min&[0,  0.00201]& 
 \makecell{$[$0,  0.00525$]$}  & \makecell{$[$0,  0.005$]$} & 
\makecell{$[$0,  0.00483$]$}  & \makecell{$[$0,  0.00955$]$}  & \makecell{$[$0.00005,  0.00959$]$}  & \makecell{$[$0.00005,  0.00959$]$}  & \makecell{$[$0.00131,  0.00959$]$} & [0,0.005]\\ 
                                  \hline
          \multirow{2}{*}{\makecell{Collision \\ (Extended)}} & max&[0.35751,  0.49961]& 
\makecell{$[$0.42267,  0.43675$]$}  & \makecell{$[$0.42418,  0.4342$]$} & \makecell{$[$0.42187,  0.43187$]$}  & \makecell{$[$0.42345,  0.43345$]$}  & \makecell{$[$0.42463,  0.43463$]$}  & \makecell{$[$0.42457,  0.43457$]$}  & \makecell{$[$0.42385,  0.43385$]$} & \makecell{$[$0.42342,  0.43342$]$}\\ 
                             & min&[0.04296,  0.06311]& \makecell{$[$0.0482,  0.0582$]$}  & \makecell{$[$0.04772,  0.05772$]$} & \makecell{$[$0.04785,  0.05785$]$}  & \makecell{$[$0.04823,  0.05823$]$}  & \makecell{$[$0.04812,  0.05812$]$}  & \makecell{$[$0.0481,  0.0581$]$}  & \makecell{$[$0.04757,  0.05772$]$}& \makecell{$[$0.04618,  0.05618$]$}\\  
                                  \hline
         \multirow{2}{*}{\makecell{Collision \\ (Advanced)}} & max&[0.14807,  0.31121]& 
\makecell{$[$0.2072,  0.2172$]$}  & \makecell{$[$0.20873,  0.21872$]$} & \makecell{$[$0.21872,  0.2185$]$}  & \makecell{$[$0.20854,  0.21854$]$}  & \makecell{$[$0.20854,  0.21854$]$}  & \makecell{$[$0.20855,  0.21855$]$}  & \makecell{$[$0.20111,  0.21111$]$}& \makecell{$[$0.20167,  0.21166$]$}\\   
                            & min&[0.02471,  0.05191]&   \makecell{$[$0.02631,  0.03631$]$}  & \makecell{$[$0.03045,  0.04045$]$} & \makecell{$[$0.03016,  0.04016$]$}  & \makecell{$[$0.03001,  0.04$]$}  & \makecell{$[$0.03001,  0.04$]$}  & \makecell{$[$0.03016,  0.04016$]$}  & \makecell{$[$0.03164,  0.04164$]$} & \makecell{$[$0.0304,  0.0404$]$}\\  
                                  \hline  
          Anesthesia & n/a &[0.00916, 0.04222]&
\makecell{$[$0.01361,  0.02361$]$}  & \makecell{$[$0.01339,  0.02332$]$} & \makecell{$[$0.01374,  0.02374$]$}  & \makecell{$[$0.01373,  0.02373$]$}  & \makecell{$[$0.01318,  0.02318$]$}  & \makecell{$[$0.01311,  0.02311$]$}  & \makecell{$[$0.01592,  0.02592$]$} & \makecell{$[$0.01815,  0.02815$]$}\\ 
          
\specialrule{.12em}{.05em}{.05em}    
        \multicolumn{11}{|c|}{\textbf{\textit{Confidence level c=0.99999}}}\\ 
    	\specialrule{.12em}{.05em}{.05em}
         \textbf{Model}  & \textbf{Type} & \textbf{\textit{P}} & \textbf{$CI_B$} & \textbf{$CI_{CLT}$}  & \textbf{$CI_{AC_W}$}& \textbf{$CI_W$} & \textbf{$CI_L$}& \textbf{$CI_{Ans}$}& \textbf{$CI_{Arc}$} & \textit{Qint} \\
			\specialrule{.1em}{.05em}{.05em}
            
      \multirow{2}{*}{Good} & max & 0.1 & \makecell{$[$0.09499,  0.10499$]$}  & \makecell{$[$0.09378,  0.10378$]$} & \makecell{$[$0.09386,  0.10386$]$}  & \makecell{$[$0.09389,  0.10389$]$}  & \makecell{$[$0.09391,  0.10391$]$}  & \makecell{$[$0.09392,  0.10392$]$}  & \makecell{$[$0.09405,  0.10405$]$} & \makecell{$[$0.09512,  0.10512$]$}\\  
         & min  & 0.1 & \makecell{$[$0.09419,  0.10419$]$}  & \makecell{$[$0.09667,  0.10667$]$} & \makecell{$[$0.09668,  0.10668$]$}  & \makecell{$[$0.09677,  0.10677$]$}  & \makecell{$[$0.09671,  0.10671$]$}  & \makecell{$[$0.09679,  0.10679$]$}  & \makecell{$[$0.09675,  0.10675$]$}  & \makecell{$[$0.09525,  0.10525$]$}\\  
                               \hline
 \multirow{3}{*}{Bad} & max & 0.95001& \makecell{$[$0.94525,  0.95525$]$}  & \makecell{$[$0.94579,  0.95579$]$} & \makecell{$[$0.94564,  0.95564$]$}  & \makecell{$[$0.94548,  0.95548$]$}  & \makecell{$[$0.94545,  0.95545$]$}  & \makecell{$[$0.94543,  0.95543$]$}  & \makecell{$[$0.94735,  0.95735$]$}  & \makecell{$[$0.94543,  0.95543$]$}\\ 
              & max2   &0.88747& \makecell{$[$0.88215,  0.89215$]$}  & \makecell{$[$0.88055,  0.89055$]$} & \makecell{$[$0.88057,  0.89057$]$}  & \makecell{$[$0.88046,  0.89046$]$}  & \makecell{$[$0.88046,  0.89046$]$}  & \makecell{$[$0.88046,  0.89046$]$}  & \makecell{$[$0.88325,  0.89325$]$}  & \makecell{$[$0.88052,  0.89052$]$}\\                     & min  & $4 \times 10^{-7}$& 
\makecell{$[$0,  0.00517$]$}  & \makecell{$[$0,  0.00319$]$} & \makecell{$[$0,  0.00494$]$}  & \makecell{$[$0,  0.00984$]$}  & \makecell{$[$0,  0.00992$]$}  & \makecell{$[$0,  0.00992$]$}  & \makecell{$[$0.00445,  0.0139$]$}  & [0,0.005]\\ 
                               \hline
         \multirow{2}{*}{\makecell{Deceleration}} & max & [0.08404,  0.08881]& 
\makecell{$[$0.08613,  0.09613$]$}  & \makecell{$[$0.08624,  0.09624$]$} & \makecell{$[$0.08312,  0.09312$]$}  & \makecell{$[$0.08725,  0.09725$]$}  & \makecell{$[$0.08725,  0.09725$]$}  & \makecell{$[$0.08726,  0.09726$]$}  & \makecell{$[$0.08746,  0.09746$]$}  & \makecell{$[$0.08737,  0.09735$]$}\\ 
                & min   & [0.04085,  0.04275] & \makecell{$[$0.03514,  0.04514$]$}  & \makecell{$[$0.03919,  0.04919$]$} & \makecell{$[$0.03918,  0.04918$]$}  & \makecell{$[$0.03942,  0.04942$]$}  & \makecell{$[$0.03943,  0.04943$]$}  & \makecell{$[$0.03944,  0.04944$]$}  & \makecell{$[$0.039,  0.049$]$}  & \makecell{$[$0.03377,  0.04377$]$} \\  
                                  \hline
         \multirow{2}{*}{\makecell{Collision \\ (Basic)}} & max& [0.96567,  0.97254]& \makecell{$[$0.96359,  0.97359$]$}  & \makecell{$[$0.96241,  0.97241$]$} & \makecell{$[$0.96767,  0.9767$]$}  & \makecell{$[$0.96892,  0.96892$]$}  & \makecell{$[$0.96689,  0.97589$]$}  & \makecell{$[$0.96683,  0.97583$]$}  & \makecell{$[$0.96863,  0.97863$]$}  & \makecell{$[$0.96462,  0.97462$]$}\\   
               & min   & [0 ,  0.00201]& \makecell{$[$0,  0.00517$]$}  & \makecell{$[$0,  0.00319$]$} & \makecell{$[$0,  0.00494$]$}  & \makecell{$[$0,  0.00984$]$}  & \makecell{$[$0,  0.00992$]$}  & \makecell{$[$0,  0.00992$]$}  & \makecell{$[$0.00445,  0.0139$]$}  & [0,0.005]\\
                                  \hline
          \multirow{2}{*}{\makecell{Collision \\ (Extended)}} & max   & [0.35751,  0.49961]&  \makecell{$[$0.42651,  0.43652$]$}  & \makecell{$[$0.42719,  0.43724$]$} & \makecell{$[$0.42757,  0.43757$]$}  & \makecell{$[$0.42656,  0.43656$]$}  & \makecell{$[$0.41774,  0.42774$]$}  & \makecell{$[$0.41779,  0.42779$]$}  & \makecell{$[$0.42745,  0.43745$]$}  & \makecell{$[$0.42875,  0.43875$]$}\\  
                 & min   &[0.04296,  0.06311]&  \makecell{$[$0.04979,  0.05979$]$}  & \makecell{$[$0.04766,  0.05766$]$} & \makecell{$[$0.04764,  0.05764$]$}  & \makecell{$[$0.04748,  0.05748$]$}  & \makecell{$[$0.04745,  0.05745$]$}  & \makecell{$[$0.04776,  0.05776$]$}  & \makecell{$[$0.05776,  0.05673$]$}  & \makecell{$[$0.04576,  0.05576$]$}\\ 
                                  \hline
         \multirow{2}{*}{\makecell{Collision \\ (Advanced)}} & max &[0.14807, 0.31121]& \makecell{$[$0.20515,  0.21519$]$}  & \makecell{$[$0.20558,  0.21563$]$} & \makecell{$[$0.20533,  0.21533$]$}  & \makecell{$[$0.20531,  0.21531$]$}  & \makecell{$[$0.20547,  0.21547$]$}  & \makecell{$[$0.20547  0.21547$]$}  & \makecell{$[$0.20385,  0.21385$]$}  & \makecell{$[$0.20453,  0.21453$]$}\\  
            & min    & [0.02471,  0.05191]  & \makecell{$[$0.03011,  0.04015$]$} & \makecell{$[$0.02902,  0.03902$]$}  & \makecell{$[$0.02954,  0.03945$]$}  & \makecell{$[$0.03956,  0.04956$]$}  & \makecell{$[$0.03861,  0.04861$]$}  & \makecell{$[$0.03887,  0.04887$]$}  & \makecell{$[$0.0363,  0.04363$]$}  & \makecell{$[$0.03031,  0.04031$]$}\\  
                                  \hline  
          Anesthesia & n/a & [0.00916,  0.04222] &  \makecell{$[$0.01284,  0.02284$]$}  & \makecell{$[$0.01513,  0.02511$]$} & \makecell{$[$0.01623,  0.02623$]$}  & \makecell{$[$0.01545,  0.02545$]$}  & \makecell{$[$0.01557,  0.02557$]$}  & \makecell{$[$0.01562,  0.02562$]$}  & \makecell{$[$0.01385,  0.02385$]$}  & \makecell{$[$0.01852,  0.02852$]$}\\
          
			\specialrule{.12em}{.05em}{.05em}
		\end{tabular}
	 \end{center}
      \end{landscape}
    \end{adjustwidth}
       \caption{Results for CI computation obtained via ProbReach, with solver precision  \textbf{$\delta$}=$10^{-3}$ and interval size  equal to $10^{-2}$,  \textbf{Type} -  extremum type and \textbf{\textit{P}} - true probability value, where single point values were analytically computed and interval values are numerically guaranteed.} \label{table:ResInt1}
	\end{sidewaystable}

\begin{table}[p]  
	\begin{adjustwidth}{-.2in}{-.1in}  
        \begin{center}
         \scriptsize 
		\label{tab:simParameters}
        	\begin{tabular}{|c|c|c|c|c|c|c|c|c|c|c|c|}
			\specialrule{.12em}{.05em}{.05em}
			 \textbf{Model}  & \textbf{Type}   & \textbf{\textit{c}} & \textbf{$CI_B$} & \textbf{$CI_{CLT}$}  & \textbf{$CI_{AC_W}$}& \textbf{$CI_W$} & \textbf{$CI_L$}& \textbf{$CI_{Ans}$}& \textbf{$CI_{Arc}$} & \textit{Qint}\\
		\specialrule{.12em}{.05em}{.05em}
         \multirow{2}{*}{Good} & max   & 0.99 & 24252  & 24025  & 24038   & 24027   & 24035   & 24034  & 26681 & 23136\\ 
                               & min    & 0,99 & 23451  & 23248  & 24256   & 24250   & 24253   & 24252  & 26894 & 23245\\
                               \hline
         \multirow{3}{*}{Bad} & max  & 0.99 & 13118  & 12670  & 12841   & 12817   & 12833   & 12832  & 23006 & 11726\\
                              & max2    &0.99 & 27498  & 26954  & 26960   & 26955   & 26958   & 26958  & 40442 & 25734\\                                  & min    & 0.99 & 2590  & 96  & 961   & 688   & 680   & 680  & 347 & 94\\
                               \hline
         \multirow{2}{*}{\makecell{Deceleration}} & max    & 0.99 & 22842  & 22393  & 22673   & 22517   & 22628   & 22623  & 24365 & 20318\\
                & min    & 0.99 & 11224  & 11073  & 11114   & 11086 & 11104   & 11104  & 11570 & 9798\\
                                  \hline
         \multirow{2}{*}{\makecell{Collision \\ (Basic)}} & max   & 0.99 & 9581  & 9318  & 9653   & 9463   & 9386   & 9381  & 10643 & 8222\\ 
                    & min   & 0.99 & 2590  & 96  & 961   & 688   & 680   & 680  & 347 & 94\\
                                  \hline
          \multirow{2}{*}{\makecell{Collision \\ (Extended)}} & max   & 0.99& 65109  & 64804  & 64854   & 64841   & 64932   & 64930  & 104637 & 62485\\ 
                             & min   & 0.99& 13624  & 13257  & 13486   & 13375   & 13326   & 13320  & 14737 & 12869\\
                                  \hline
         \multirow{2}{*}{\makecell{Collision \\ (Advanced)}} & max   & 0.99 & 44370  & 43602  & 43645   & 43640   & 43644   & 43643  & 51734 & 43524\\  
                            & min     & 0.99 & 9500  & 9081  & 9094   & 9085   & 9090   & 9089  & 9282 & 9080\\ 
                                  \hline  
          Anesthesia & n/a  & 0.99 & 5801  & 4847  & 5024   & 4952   & 4928   & 4919  & 5522 & 4804\\
			
            \specialrule{.12em}{.05em}{.05em}
             
                \multirow{2}{*}{Good} & max  & 0.99999 & 70422  & 69484  & 69582   & 69496   & 69530   & 69529  & 77262 & 68456\\ 
                               & min   & 0,99999 & 71898  & 71286  & 71339   & 71293   & 71321  & 71321  & 79369 & 68994\\
                               \hline
         \multirow{3}{*}{Bad} & max  & 0.99999 & 37388  & 36518  & 36771   & 36629   & 36687  & 36868 & 60006 & 36164\\
                              & max2     &0.99999 & 79306  & 79097  & 79125   & 79101  & 79118  & 79118  & 96442 & 77892\\                                  & min    & 0.99999 & 5797  & 124  & 2766   & 1963   & 4136   & 4136  & 572 & 94\\
                               \hline
         \multirow{2}{*}{\makecell{Deceleration}} & max   & 0.99999 & 65248  & 65233  & 65330   & 65299   & 65320   & 65319  & 72114 & 59882\\
                & min   & 0.99999 & 33147  & 32969  & 33133   & 33018& 33060  & 33060  & 34231 & 29096\\
                                  \hline
         \multirow{2}{*}{\makecell{Collision \\ (Basic)}} & max  & 0.99999 & 25279  & 24711  & 24834   & 24789   & 24934  & 24933  & 26045 & 23016\\ 
                    & min   & 0.99999& 5797  & 124  & 2766   & 1963   & 4136   & 4136  & 572 & 94\\
                                  \hline
          \multirow{2}{*}{\makecell{Collision \\ (Extended)}} & max   & 0.99999& 191466  & 190776  & 191253   & 190894   & 191485   & 191472  & 376294 & 185456\\ 
                             & min   & 0.99999& 41153  & 38942  & 39745   & 39473  & 39537   & 39541  & 47923 & 37608\\
                                  \hline
         \multirow{2}{*}{\makecell{Collision \\ (Advanced)}} & max  & 0.99999 & 131517 & 129746  & 131185  & 129845  & 129934  & 129933  & 183405 & 127486\\  
                            & min    & 0.99999 & 27305  & 25657  & 25835  & 25736  & 25792  & 25791  & 29362 & 24569\\ 
                                  \hline  
          Anesthesia & n/a  & 0.99999 & 16197  & 15453 & 15834  & 15634  & 15734  & 15733 & 17845 & 15314\\
          
		\specialrule{.12em}{.05em}{.05em} 
		\end{tabular}
	 \end{center}
    \end{adjustwidth}
      \caption{Samples size comparison for confidence interval computation obtained via ProbReach, with solver  \textbf{$\delta$} precision equal to $10^{-3}$ and interval size equal to $10^{-2}$,  \textbf{Type} -  extremum type and  \textbf{\textit{c}} - confidence level.} \label{table:ResSam1}
	\end{table}
    
\subsection{Tested Models Results} \label{sec:Res-models}

\paragraph{Intervals based on CLT and Bayesian interval.}
Based on our model set, we provide in Table \ref{table:ResInt1} a comparison of the CIs described in the Section \ref{sec:CI}, obtained via ProbReach with precision $\delta= 10^{-3}$, interval size $10^{-2}$ and true probability value \textbf{\textit{P}}is either analytically computed single probability values or formally computed absolute (non-statistical) intervals. These parameters were chosen according to previous work \cite{19} as a fine trade-off between the precision of the results and the CPU time. Each model was verified separately with different confidence level from 0.99 to 0.99999 (see also Table \ref{table:ResInt2}). The lowest confidence level is often used in similar work, while the highest can provide reasonable results for real-world complex models.

As it can be seen in Table \ref{table:ResInt1}, all the intervals for the various techniques overlap. The modified $CI_{CLT}$ approach shows very similar results to the $CI_{B}$, which can be regarded as a successful implementation. The key difference in the interval sizes can be found in the results of the \say{Bad} model Type min and the Collision (Basic) model Type min. From the results we can conclude that the true probability value is very close to 0. This allows  the Bayesian, CLT and Agresti-Coull methods to form intervals, which in reality are half of the proposed interval size $10^{-2}$, while the other techniques return \say{fully} sized intervals. That happens because $CI_{B}$ is using posterior distribution to form the interval. At the same time, the $CI_{CLT}$ and $CI_{AC_W}$ calculations of the mean value return the result, which is quite close to zero. Thus, the next step of the interval bounds computation cuts the negative part of the interval. This trend is holding for all probability values within $[0,0.001]$.

Table \ref{table:ResInt1} also shows that with the increase of the confidence level the interval's precision is growing, which in turn is directly related to the usage of the inverse cumulative distribution function for normal random variable with given confidence level in formulas for $CI_{CLT}$ (\ref{CI:CLT}), $CI_{W}$ (\ref{CI:Wilson}), $CI_{AC_{W}}$ (\ref{CI:AC_W}) and $CI_{Arc}$ (\ref{CI:Arcsine}). It also results in the increase of the sample size $n$ for $CI_{L}$ and $CI_{Anc}$.

The comparison of the obtained intervals (see Table \ref{table:ResInt1}) with the true probability value or interval \textbf{\textit{P}} shows that all CIs contain the single probability values, but $CI_{Acr}$ (see  \say{Bad} type min model of Table \ref{table:ResInt1}), and all CIs overlap with the true probability intervals. We can also note that the true probability intervals of the Collision Extended, Collision Advanced, and Anesthesia models contain all confidence intervals for all confidence levels (see also Table \ref{table:ResInt2}). The reasons why Collision Basic and Deceleration models' true probability intervals do not contain CIs are their size, which is very small ($<0.01$) and their closeness to 0.

Table \ref{table:ResSam1} provides very interesting results with respect to the number of samples, which were used to find CIs obtained via ProbReach with solver precision $\delta=10^{-3}$ and interval size $10^{-2}$. The number of samples varies for different models and types (see also Table \ref{table:ResSam2}). As it was noted earlier for Figure 4, the number of samples needed for the computation grows from the bounds to the center of the [0,1] interval. The presented models show different behaviour and probability results. The most important outcome is that all CIs (except $CI_{Arc}$) show better result in number of points with respect to $CI_{B}$. The best result was shown by $CI_{CLT}$. It shows that the proposed CLT modification can provide reasonable results for RQMC calculation in comparison with the well-established Bayesian MC integral calculation.


\paragraph{Qint method results.}


A comparison of the Qint method's confidence intervals is presented in Table \ref{table:ResInt1}. All of the Qint intervals also contain single probability values and overlap with true probability intervals. The original Qint algorithm is not able to provide results for \say{Bad} type min and Collision Basic type min models, because for very small probability values like $4 \times 10^{-7}$ and [0, 0.00201] it could not detect $n_s > 0$ for the chosen confidence levels and interval size. Due to this reason the original Qint algorithm was changed by modifying the CLT method described in 
Section \ref{sec:CLT-CI}. From the results we see that that the Qint algorithm shows great potential, which is connected with the very fast convergence rate of the QMC method and with finding an appropriate partition (in terms of the parameters $k,s$).

Table \ref{table:ResSam1} allows us to compare Qint's sample sizes with those of other CIs. However, $CI_{CLT}$ had an advantage in the number of samples for small probability values near the border (see Figure \ref{fig:Points}), where we can clearly see that for bigger true probability values, which is presented in the tested models except \say{Bad} type min and Collision Basic type min, this trend is not preserved. On the contrary,  Qint uses fewer number of samples than  other CIs and $CI_{CLT}$ in particular (see also Figure \ref{fig:Probability}). For the tested models set with confidence \textbf{\textit{c}}=0.99999, Qint used on average between 1,850 and 24,802 fewer samples than other CIs techniques.

\section{Conclusion and Future Work}

In this paper we provide comprehensive evaluation of CIs calculation techniques based on Monte Carlo (MC) and Quasi-Monte Carlo (QMC) techniques. The experiments show that our modified CLT technique is usable in practice even for complex dynamics and for probabilities close to the bounds. The QMC-based calculation techniques we consider have excellent convergence and efficiency especially when the number of samples is small. Based on our analysis of CIs, we suggest that our results can be used as guidelines for probability estimation techniques. In the future we plan to provide specification for the CI calculation by the Qint technique for the chosen models and also extend the test model range. Possible variations of the Qint technique are also of interest.  In particular, joining the CLT and Qint methods is the next challenge for further research.

\bibliographystyle{eptcs}

\appendix
\section{Appendix}

\subsection{Formal definition of PHS \cite{20}}\label{appendix:PHS}

A Parametric Hybrid System (PHS) is a tuple
\[
H=<Q,\Upsilon,X,P,Y,R,\textnormal{jump, goal}>
\]
where
\begin{itemize}
\item $Q = \{q_0, \cdots , q_m\}$ a set of modes (discrete components of the system),
\item $\Upsilon = \{(q,q^{\prime}): q,q^{\prime} \in Q\}$ a set of transitions between modes,
\item $X = [u_1, v_1] \times \cdots \times [u_n, v_n] \subset \mathbb{R}^n$ a domain of continuous variables,
\item $P = [a_{1}, b_{1}] \times \cdots \times [a_{k}, b_{k}] \subset \mathbb{R}^k$ the parameter space of the system,
\item $Y = \{{{\bf y}_{q}({\bf p}, t)} : q \in Q, {\bf p}\in X\times P, t\in[0,T]\}$ the continuous system dynamics where ${\bf y}_q:X\times P\times [0, T] \rightarrow X$,
\item $R = \{{\bf g}_{(q,q^{\prime})}({\bf p},t) : (q, q^{\prime}) \in \Upsilon, {\bf p}\in X\times P, t\in[0,T]\}$ `reset' functions ${\bf g}_{(q,q^{\prime})}: X\times P\times [0,T] \rightarrow X$ defining the continuous state at time $t = 0$ in mode $q^{\prime}$ after taking the transition from mode $q$.
\end{itemize}
and predicates (or relations)
\begin{itemize}
        \item $\textnormal{jump}_{(q, q^\prime)}({\bf x})$ defines a discrete transition $(q, q^{\prime}) \in \Upsilon$ which may (but does not have to) occur upon reaching the jump condition in state $({\bf x},q)\in X\times P\times Q$,
        \item $\textnormal{goal}_{q}({\bf x})$ defines the goal state ${\bf x}$ in mode $q$.
\end{itemize}

\subsection{Koksma-Hlawka inequality}\label{appendix:KH}

The well-known Koksma-Hlawka inequality \cite{17} provides an upper bound for the integral estimation error with QMC methods. Suppose we want to compute $I=\int_{U_d}f(x)dx$, where $U_d$ is the hypercube over $[0,1]^d$.
 Let $\{u_1, ..., u_n\}$ be a set in $U_d$. Then the Koksma-Hlawka inequality is:
\begin{equation} \label{eq:KH}
\bigg| I-\frac{1}{n}\sum_{i=1}^n f(u_i)\bigg| \leqslant V(f)D^*_n \{u_1, ..., u_n\},
\end{equation}
where $V(f)$ is the bounded variation in the sense of Hardy and Krause:
$$
V(f)=\sum_{k=1}^d \sum_{1<i_1<...<i_k<d} V_{V_{it}}^k(f;i_1,...,i_k),
$$
where $V_{V_{it}}^k(f;i_1,...,i_k)$ is the variation in sense of Vitali \cite{23}, applied to the restriction of $f$ to the space dimension $k\{(u_1,...,u_d)\in [0,1]^d: u_j=1$ for $j\neq i_1,...,i_k\}$. If $k=d$ we obtain an empty set, which can not be calculated.


The star-discrepancy $D^*_n$ is defined as follows:
$$
\mathbb{D}^*_n\{u_1, ..., u_n\}=\underset{B\in W^*}{sup} \bigg|\frac{\#\{u_i:u_i\in B\}}{n}-\lambda_d(B)\bigg|,
$$
where $\#\{u_i:u_i\in B\}$ are points from the set $B$ and $W^*$ is defined as the set of the form:
$$
\prod_{k=1}^d[0,c_k ) =\{y\in U_d:0 \leqslant y_k < c_k\}
$$

 Unfortunately inequality (\ref{eq:KH}) can not serve as a basis for a constructive evaluation of the integration error in practical applications. In particular, computing the star-discrepancy of an arbitrary set is an NP-hard problem \cite{15}. Also, estimating the Hardy-Krause variation is a very complicated computational problem.


\subsection{Software} \label{appendix:Soft}
We conducted our research on the basis of the ProbReach tool \cite{19} for computing bounded reachability in stochastic parametric hybrid systems. ProbReach can use either dReal \cite{13} or iSAT-ODE \cite{10} solvers for analyzing (standard) bounded reachability question. 
Essentially, both dReal and iSAT-ODE are implementations of a $\delta$-complete decision procedure for first-order logic functions over the reals that may include nonlinear functions and ordinary differentiation equations. We performed our research by using dReal3 tool (version 3.16.08.01).






\subsection{Tested models} \label{appendix:Models}
The main aim of the tests conducted is to reveal how different model types and their complexity can affect the computational result. 
Five different hybrid models were chosen for our experiments can be found at: https://github.com/dreal/probreach/tree/master/model.

\paragraph{ ``Good'' and ``Bad'' models.}
These models represent single mode non-hybrid systems with constant flow dynamics. Initially, both systems have a state defined by the predicate $(x(0)=r)\wedge (n\in [0,1])$, where $n$ is a  parameter on $[0,1]$ and $r$ is uniformly distributed over [0,1]. The probability function in the \say{Good} model case is constant $p(n)=0.1$, while in the \say{Bad} model case it is equal to $p(n)=4n^2-4n+1$, which reaches its minimum value of 0 at $n=0.5$ and the maximum value of 1 at $n=0$  and $n=1$.
ProbReach is used to compute 0-step bounded reachability probability for the \say{Good} model goal specified by the predicate $(x\leq 0.9n+0.1) \wedge (x \geq 0.9n)$ and for the \say{Bad} model goal specified by the predicate $(x\leq 2(n+0.5)^2+0.5) \wedge (x \geq -2(n-0.5)^2+0.5).$

\paragraph{Deceleration model.}
This hybrid model represents a car deceleration scenario. During the first mode the car accelerates from 0 to 100 km/h, using a nonlinear function that depends on random parameter. During the second mode driver is reacting after 1.2 seconds and starting deceleration. The final mode presents car deceleration. The model takes into account that velocity changes due to friction. ProbReach is used to compute the probability of a car stopping within 400 meters. 

\paragraph{Collision model.}
This hybrid model represents a two car collision scenario. Car1 and Car2 move on the same lane with speed 11.12 $m/sec$. During the first mode Car2 continues movement, while Car1 changes lane and starts accelerating until it gets ahead of Car2 by a safe distance. During the second mode Car2 returns to the initial lane and starts decelerating, while Car2's driver takes time to react and then starts deceleration. ProbReach is used to compute the probability of the cars colliding for three different configurations of this model: the $basic$ model contains one random parameter - deceleration of Car1 ; $extended$ and $advanced$ models contains two random parameters for the deceleration of Car1 and Car2.

\paragraph{Anesthesia model.}
This model represents anesthesia delivery. It tracks the drug concentration which changes as the result of metabolic processes in human body. The model's scenario assumes that drug delivery is continuous, but every 15 minutes (starting at time 0) the drug infusion rate is subject to random errors. ProbReach computes the probability of reaching the unsafe state $(c_p(t)\geq 6) \vee (c_p(t) \leq 1) \vee (c_1(t)\geq 10) \vee(c_1(t) \leq0) \vee(c_2(t)\geq 10) \vee (c_2(t) \leq 0)$ in 3 jumps within 60 minutes, where $c_p$ - plasma drug concentration, $c_1$- fast peripheral compartment drug concentration and $c_2$ - slow peripheral compartment drug concentration.    
$$
$$

\subsection{Further Results}
$$
$$
$$
$$
$$
$$
$$
$$

\begin{sidewaystable}[p] 
	\begin{adjustwidth}{-.3in}{-.3in}  
     \begin{landscape}
        \begin{center}
        \scriptsize 
		\label{tab:simParameters}
		\begin{tabular}{|c|c|c|c|c|c|c|c|c|c|c|c|}

		      \specialrule{.12em}{.05em}{.05em}    
        \multicolumn{11}{|c|}{\textbf{\textit{Confidence level c=0.999}}}\\ 
    	\specialrule{.12em}{.05em}{.05em}
			 \textbf{Model}  & \textbf{Type} & \textbf{\textit{P}} & \textbf{$CI_B$} & \textbf{$CI_{CLT}$}  & \textbf{$CI_{AC_W}$}& \textbf{$CI_W$} & \textbf{$CI_L$}& \textbf{$CI_{Ans}$}& \textbf{$CI_{Arc}$} & \textit{Qint}\\
			\specialrule{.12em}{.05em}{.05em}

         \multirow{2}{*}{Good} & max&0.1& \makecell{$[$0.09555,  0.10555$]$}  & \makecell{$[$0.09559,  0.10559$]$} & \makecell{$[$0.09559,  0.10559$]$}  & \makecell{$[$0.0957,  0.1057$]$}  & \makecell{$[$0.09572,  0.10572$]$}  & \makecell{$[$0.09571,  0.10571$]$}  & \makecell{$[$0.09528,  0.10528$]$}  & \makecell{$[$0.09444,  0.10444$]$}\\  
                               & min&0.1& \makecell{$[$0.09393, 0.10393$]$}  & \makecell{$[$0.0961,  0.1061$]$} & \makecell{$[$0.09613,  0.10613$]$}  & \makecell{$[$0.0962,   0.1062$]$}  & \makecell{$[$0.09619,  0.10619$]$}  & \makecell{$[$0.0962,  0.1062$]$}  & \makecell{$[$0.09637,  0.10637$]$}  & \makecell{$[$0.09263,  0.10263$]$}\\ 
                               \hline
         \multirow{3}{*}{Bad} & max&0.95001& \makecell{$[$0.94549,  0.95549$]$}  & \makecell{$[$0.94544,  0.95544$]$} & \makecell{$[$0.94526,  0.95526$]$}  & \makecell{$[$0.94504,  0.95504$]$}  & \makecell{$[$0.94502,  0.95502$]$}  & \makecell{$[$0.94499,  0.95499$]$}  & \makecell{$[$0.94735,  0.95735$]$}  & \makecell{$[$0.94564,  0.95564$]$}\\  
                              & max2&0.88747& \makecell{$[$0.88165,  0.89165$]$}  & \makecell{$[$0.88069,  0.89069$]$} & \makecell{$[$0.88071,  0.89071$]$}  & \makecell{$[$0.8806,  0.8906$]$}  & \makecell{$[$0.88061,  0.89061$]$}  & \makecell{$[$0.88061,  0.89061$]$}  & \makecell{$[$0.88325,  0.89325$]$}  & \makecell{$[$0.88059,  0.89059$]$}\\                                   & min&$4 \times 10^{-7}$& 
 \makecell{$[$0,  0.00525$]$}  & 
\makecell{$[$0,  0.005$]$} & 
\makecell{$[$0,  0.00489$]$}  & 
\makecell{$[$0,  0.00972$]$}  & \makecell{$[$0.00005,  0.00978$]$}  & \makecell{$[$0.00005,  0.00978$]$}  & \makecell{$[$0.00024,  0.01173$]$}  & [0,0.005]\\ 
                               \hline
         \multirow{2}{*}{\makecell{Deceleration}} & max&[0.08404,  0.08881]& 
\makecell{$[$0.08695,  0.09695$]$}  & \makecell{$[$0.08659,  0.09659$]$} & \makecell{$[$0.08656,  0.09656$]$}  & \makecell{$[$0.0867,  0.0967$]$}  & \makecell{$[$0.08675,  0.09675$]$}  & \makecell{$[$0.08683,  0.09683$]$}  & \makecell{$[$0.0868,  0.0968$]$}  & \makecell{$[$0.08825,  0.08925$]$}\\  
                & min&[0.04085,  0.04275]&  
\makecell{$[$0.03785,  0.04785$]$}  & \makecell{$[$0.0362,  0.0462$]$} & \makecell{$[$0.04,  0.05$]$}  & \makecell{$[$0.0403,  0.0503$]$}  & \makecell{$[$0.0402,  0.0502$]$}  & \makecell{$[$0.0403,  0.0503$]$}  & \makecell{$[$0.04001,  0.05001$]$}   & \makecell{$[$0.03495,  0.04495$]$} \\  
                                  \hline
         \multirow{2}{*}{\makecell{Collision \\ (Basic)}} & max&[0.96567,  0.97254]& 
\makecell{$[$0.96397,  0.97052$]$}  & \makecell{$[$0.96362,  0.97362$]$} & \makecell{$[$0.96392,  0.97392$]$}  & \makecell{$[$0.96412,  0.97412$]$}  & \makecell{$[$0.96521,  0.97521$]$}  & \makecell{$[$0.96516,  0.97516$]$}  & \makecell{$[$0.96851,  0.97851$]$}  & \makecell{$[$0.96305,  0.97305$]$}\\  
                & min&[0,  0.00201]& \makecell{$[$0,  0.00525$]$}  & 
\makecell{$[$0,  0.005$]$} & 
\makecell{$[$0,  0.00489$]$}  & 
\makecell{$[$0,  0.00972$]$}  & \makecell{$[$0.00005,  0.00978$]$}  & \makecell{$[$0.00005,  0.00978$]$}  & \makecell{$[$0.00024,  0.01173$]$}  & [0,0.005]\\
                                  \hline
          \multirow{2}{*}{\makecell{Collision \\ (Extended)}} & max&[0.35751,  0.49961]& 
\makecell{$[$0.42673,  0.43676$]$}  & \makecell{$[$0.42477,  0.43477$]$} & \makecell{$[$0.42634,  0.42634$]$}  & \makecell{$[$0.42441,  0.42441$]$}  & \makecell{$[$0.42448,  0.42448$]$}  & \makecell{$[$0.42434,  0.42434$]$}  & \makecell{$[$0.42345,  0.43345$]$}  & \makecell{$[$0.43553,  0.45553$]$}\\  
                             & min&[0.04296,  0.06311]&  \makecell{$[$0.05004,  0.06004$]$}  & \makecell{$[$0.05173,  0.06173$]$} & \makecell{$[$0.04189,  0.05189$]$}  & \makecell{$[$0.04244,  0.05244$]$}  & \makecell{$[$0.04385,  0.05385$]$}  & \makecell{$[$0.04382,  0.05382$]$}  & \makecell{$[$0.04483,  0.05483$]$}  & \makecell{$[$0.04499,  0.05499$]$}\\ 
                                  \hline
         \multirow{2}{*}{\makecell{Collision \\ (Advanced)}} & max&[0.14807, 0.31121]& 
\makecell{$[$0.20451,  0.21456$]$}  & \makecell{$[$0.20639,  0.21643$]$} & \makecell{$[$0.20623,  0.21623$]$}  & \makecell{$[$0.20617,  0.21617$]$}  & \makecell{$[$0.20647,  0.21647$]$}  & \makecell{$[$0.20643,  0.21643$]$}  & \makecell{$[$0.20111,  0.21111$]$}  & \makecell{$[$0.20469,  0.21469$]$}\\  
                            & min&[0.02471,  0.05191]& \makecell{$[$0.02626,  0.03633$]$}  & \makecell{$[$0.03256,  0.04256$]$} & \makecell{$[$0.02666,  0.03666$]$}  & \makecell{$[$0.03212,  0.04212$]$}  & \makecell{$[$0.03301,  0.04301$]$}  & \makecell{$[$0.03301,  0.04301$]$}  & \makecell{$[$0.03364,  0.04364$]$}  & \makecell{$[$0.03112,  0.04112$]$}\\  
                                  \hline  
          Anesthesia & n/a&[0.00916,  0.04222]&
\makecell{$[$0.01285,  0.02285$]$}  & \makecell{$[$0.01497,  0.02495$]$} & \makecell{$[$0.01574,  0.02574$]$}  & \makecell{$[$0.01492,  0.02492$]$}  & \makecell{$[$0.01446,  0.02446$]$}  & \makecell{$[$0.01442,  0.02442$]$}  & \makecell{$[$0.01592,  0.02592$]$}  & \makecell{$[$0.01774,  0.02774$]$}\\ 

      \specialrule{.12em}{.05em}{.05em}    
        \multicolumn{11}{|c|}{\textbf{\textit{Confidence level c=0.9999}}}\\ 
    	\specialrule{.12em}{.05em}{.05em}
    		 \textbf{Model}  & \textbf{Type} & \textbf{\textit{P}} & \textbf{$CI_B$} & \textbf{$CI_{CLT}$}  & \textbf{$CI_{AC_W}$}& \textbf{$CI_W$} & \textbf{$CI_L$}& \textbf{$CI_{Ans}$}& \textbf{$CI_{Arc}$} & \textit{Qint}\\
			\specialrule{.12em}{.05em}{.05em}
            
     \multirow{2}{*}{Good} & max &0.1& \makecell{$[$0.09621,  0.10621$]$}  & \makecell{$[$0.09465,  0.10465$]$} & \makecell{$[$0.09464,  0.10464$]$}  & \makecell{$[$0.09478,  0.10478$]$}  & \makecell{$[$0.09477,  0.10477$]$}  & \makecell{$[$0.09424,  0.10424$]$}  & \makecell{$[$0.09423,  0.10423$]$}  & \makecell{$[$0.09415,  0.10415$]$}\\ 
                & min  & 0.1 & 
\makecell{$[$0.09352,  0.10352$]$}  & \makecell{$[$0.09662,  0.10662$]$} & \makecell{$[$0.09669,  0.10669$]$}  & \makecell{$[$0.09672,  0.10672$]$}  & \makecell{$[$0.09674,  0.10674$]$}  & \makecell{$[$0.09674,  0.10674$]$}  & \makecell{$[$0.09684,  0.10684$]$}  & \makecell{$[$0.09489,  0.10489$]$}\\ 
                               \hline
    \multirow{3}{*}{Bad} & max & 0.95001& \makecell{$[$0.94477,  0.95477$]$}  & \makecell{$[$0.94595,  0.95595$]$} & \makecell{$[$0.94598,  0.95598$]$}  & \makecell{$[$0.94574,  0.95574$]$}  & \makecell{$[$0.94576,  0.95576$]$}  & \makecell{$[$0.94574,  0.95574$]$}  & \makecell{$[$0.95377,  0.96377$]$}  & \makecell{$[$0.94658,  0.95658$]$}\\ 
                       & max2 &0.88747& \makecell{$[$0.88208,  0.89208$]$}  & \makecell{$[$0.88058,  0.89058$]$} & \makecell{$[$0.88061,  0.89061$]$}  & \makecell{$[$0.88049,  0.89049$]$}  & \makecell{$[$0.88051,  0.89051$]$}  & \makecell{$[$0.88053,  0.89053$]$}  & \makecell{$[$0.88325,  0.89325$]$}  & \makecell{$[$0.88051,  0.89051$]$}\\                                   & min & $4 \times 10^{-7}$& \makecell{$[$0,  0.00525$]$}  & \makecell{$[$0,  0.005$]$} & 
\makecell{$[$0,  0.00492$]$}  & \makecell{$[$0,  0.00979$]$}  & \makecell{$[$0,  0.00987$]$}  & \makecell{$[$0,  0.00987$]$}  & \makecell{$[$0.00349,  0.01287$]$}  & [0,0.005]\\  
                               \hline
         \multirow{2}{*}{\makecell{Deceleration}} & max & [0.08404,  0.08881]& 
\makecell{$[$0.08593,  0.09593$]$}  & \makecell{$[$0.08631,  0.09631$]$} & \makecell{$[$0.08629,  0.09629$]$}  & \makecell{$[$0.08642,  0.09642$]$}  & \makecell{$[$0.08643,  0.09643$]$}  & \makecell{$[$0.08644,  0.09644$]$}  & \makecell{$[$0.08672,  0.09672$]$}  & \makecell{$[$0.08799,  0.09799$]$}\\  
                & min  & [0.04085,  0.04275] & \makecell{$[$0.03764,  0.04764$]$}  & \makecell{$[$0.0394,  0.0494$]$} & \makecell{$[$0.03438,  0.04438$]$}  & \makecell{$[$0.03972,  0.04972$]$}  & \makecell{$[$0.03969,  0.04969$]$}  & \makecell{$[$0.03971,  0.04971$]$}  & \makecell{$[$0.03938,  0.04938$]$}   & \makecell{$[$0.03359,  0.04359$]$} \\ 
                                  \hline
         \multirow{2}{*}{\makecell{Collision \\ (Basic)}} & max& [0.96567,  0.97254]& \makecell{$[$0.96371,  0.97373$]$}  & \makecell{$[$0.96285,  0.97285$]$} & \makecell{$[$0.96799,  0.97799$]$}  & \makecell{$[$0.96792,  0.97792$]$}  & \makecell{$[$0.96775,  0.97775$]$}  & \makecell{$[$0.96772,  0.97772$]$}  & \makecell{$[$0.96851,  0.97851$]$}  & \makecell{$[$0.96411,  0.97411$]$}\\  
                     & min  &  [0,  0.00201]& \makecell{$[$0,  0.00525$]$}  & \makecell{$[$0,  0.005$]$} & 
\makecell{$[$0,  0.00492$]$}  & \makecell{$[$0,  0.00979$]$}  & \makecell{$[$0,  0.00987$]$}  & \makecell{$[$0,  0.00987$]$}  & \makecell{$[$0.00349,  0.01287$]$}  & [0,0.005]\\ 
                                  \hline
          \multirow{2}{*}{\makecell{Collision \\ (Extended)}} & max& [0.35751,  0.49961]&  \makecell{$[$0.42492,  0.43495$]$}  & \makecell{$[$0.42799,  0.43804$]$} & \makecell{$[$0.42157,  0.43157$]$}  & \makecell{$[$0.42356,  0.43356$]$}  & \makecell{$[$0.4177,  0.42783$]$}  & \makecell{$[$0.42187,  0.43187$]$}  & \makecell{$[$0.42345,  0.43345$]$}  & \makecell{$[$0.42656,  0.43656$]$}\\ 
                   & min &[0.04296, 0.06311]&  \makecell{$[$0.04933,  0.05933$]$}  & \makecell{$[$0.04708,  0.05708$]$} & \makecell{$[$0.04864,  0.05864$]$}  & \makecell{$[$0.04848,  0.05848$]$}  & \makecell{$[$0.04845,  0.05845$]$}  & \makecell{$[$0.04845,  0.05845$]$}  & \makecell{$[$0.04254,  0.05254$]$}  & \makecell{$[$0.04463,  0.05463$]$}\\ 
                                  \hline
         \multirow{2}{*}{\makecell{Collision \\ (Advanced)}} & max &[0.14807,  0.31121]& \makecell{$[$0.20531,  0.21537$]$}  & \makecell{$[$0.20617,  0.21621$]$} & \makecell{$[$0.20633,  0.21633$]$}  & \makecell{$[$0.20631,  0.21631$]$}  & \makecell{$[$0.20647,  0.21647$]$}  & \makecell{$[$0.20646,  0.21646$]$}  & \makecell{$[$0.20834,  0.21834$]$}  & \makecell{$[$0.20496,  0.21496$]$}\\  
                 & min  & [0.02471,  0.05191] &  \makecell{$[$0.02895,  0.03895$]$}  & \makecell{$[$0.02937,  0.03937$]$} & \makecell{$[$0.02984,  0.03984$]$}  & \makecell{$[$0.03823,  0.04823$]$}  & \makecell{$[$0.03971,  0.04971$]$}  & \makecell{$[$0.03971,  0.04971$]$}  & \makecell{$[$0.03473,  0.04473$]$}  & \makecell{$[$0.03062,  0.04062$]$}\\  
                                  \hline  
          Anesthesia & n/a&[0.00916,  0.04222] &  \makecell{$[$0.01388,  0.02388$]$}  & \makecell{$[$0.01428,  0.02427$]$} & \makecell{$[$0.01399,  0.02399$]$}  & \makecell{$[$0.01425,  0.02425$]$}  & \makecell{$[$0.01483,  0.02483$]$}  & \makecell{$[$0.01493,  0.02493$]$}  & \makecell{$[$0.01623,  0.02623$]$}  & \makecell{$[$0.01847,  0.0284$]$}\\

			\specialrule{.12em}{.05em}{.05em}
		\end{tabular}
	 \end{center}
      \end{landscape}
    \end{adjustwidth}
       \caption{Comparison of results for confidence interval computation obtained via ProbReach, with solver  \textbf{$\delta$} precision equal to $10^{-3}$ and interval size  equal to $10^{-2}$,  \textbf{Type} -  extremum type and \textbf{\textit{P}} - true probability value, where single point values were analytically computed and interval values are numerically guaranteed.} \label{table:ResInt2}
	\end{sidewaystable}

\begin{table}[p] 
	\begin{adjustwidth}{-.2in}{-.1in}  
        \begin{center}
         \scriptsize 
		\label{tab:simParameters}
        	\begin{tabular}{|c|c|c|c|c|c|c|c|c|c|c|c|}
			   \specialrule{.12em}{.05em}{.05em}
			 \textbf{Model}  & \textbf{Type}   & \textbf{\textit{c}} & \textbf{$CI_B$} & \textbf{$CI_{CLT}$}  & \textbf{$CI_{AC_W}$}& \textbf{$CI_W$} & \textbf{$CI_L$}& \textbf{$CI_{Ans}$}& \textbf{$CI_{Arc}$} & \textit{Qint}\\

            \specialrule{.12em}{.05em}{.05em}
            
               \multirow{2}{*}{Good} & max  & 0.999 & 39211  & 39187  & 39215   & 39196   & 39210   & 39200  & 43407 & 38094\\ 
                               & min    & 0,999 & 39650  & 39364  & 39401   & 39368   & 39373   & 39373  & 43848 & 38204\\
                               \hline
         \multirow{3}{*}{Bad} & max  & 0.999 & 20717  & 20401  & 20550   & 20497   & 20527   & 20562  & 32006 & 20322\\
                              & max2   &0.999 & 44557  & 43848  & 43863   & 43848   & 43855   & 43855  & 56442 & 42888\\                                  & min   & 0.999 & 3950  & 107  & 1549   & 1103   & 1362   & 1362  & 434 & n/a\\
                               \hline
         \multirow{2}{*}{\makecell{Deceleration}} & max   & 0.999 & 36609  & 36039  & 36061   & 36044   & 36132   & 36130  & 39524 & 33068\\
                & min   & 0.999 & 18727  & 18629  & 18709   & 18671   & 18628   & 18682  & 19438 & 16618\\
                                  \hline
         \multirow{2}{*}{\makecell{Collision \\ (Basic)}} & max  & 0.999 & 13795  & 13222  & 13341   & 13286   & 13311   & 13397  & 15385 & 13098\\ 
                    & min  & 0.999 & 3950  & 107  & 1549   & 1103   & 1362   & 1362  & 434 & n/a\\
                                  \hline
          \multirow{2}{*}{\makecell{Collision \\ (Extended)}} & max  & 0.999& 106252  & 106099  & 106243   & 106147   & 106224   & 106224  & 166345 & 104531\\ 
                             & min  & 0.999& 22887  & 21860  & 22196   & 21935   & 22041   & 22038  & 24742 & 20862\\
                                  \hline
         \multirow{2}{*}{\makecell{Collision \\ (Advanced)}} & max  & 0.999 & 71746  & 70435  & 70646   & 70636   & 70642   & 70640  & 143390 & 69642\\  
                            & min   & 0.999 & 15833  & 15679  & 15746  & 15723  & 15748   & 15746  & 18354 & 15086\\ 
                                  \hline  
          Anesthesia & n/a & 0.999 & 9017  & 8516  & 8827   & 8628   & 8593   & 8592  & 9284 & 8430\\
			
             \specialrule{.12em}{.05em}{.05em}
            
                   \multirow{2}{*}{Good} & max   & 0.9999 & 55187  & 54327  & 54361   & 54347   & 54355   & 54362  & 60104 & 52990\\ 
                               & min   & 0,9999 & 55885  & 55281  & 55231   & 55286   & 55307   & 55307  & 61631 & 53411\\
                               \hline
         \multirow{3}{*}{Bad} & max   & 0.9999 & 29147  & 28240  & 28339  & 28276   & 28289   & 28289  & 42463 & 27944\\
                              & max2   &0.9999 & 62735  & 61139  & 61364   & 61152   & 61359   & 61358  & 86442 & 59012\\                                  & min & 0.9999 & 4849  & 116  & 2153   & 1530   & 2458   & 2458  & 508 & n/a\\
                               \hline
         \multirow{2}{*}{\makecell{Deceleration}} & max   & 0.9999 & 50476  & 50243  & 50277   & 50250   & 50269   & 50268  & 55495 & 46084\\
                & min   & 0.9999 & 25741  & 25695  & 25817   & 25779   & 25794   & 25794  & 26790 & 22466\\
                                  \hline
         \multirow{2}{*}{\makecell{Collision \\ (Basic)}} & max   & 0.9999 & 19476  & 18907  & 19084   & 18984   & 19035   & 19032  & 21537 & 18128\\ 
                    & min   & 0.9999 & 4849  & 116  & 2153   & 1530   & 2458   & 2458  & 508 & n/a\\
                                  \hline
          \multirow{2}{*}{\makecell{Collision \\ (Extended)}} & max  & 0.9999& 148388  & 147675  & 147834  & 147746  & 147786  & 147635  & 236423 & 145974\\ 
                             & min   & 0.9999& 31528  & 29894  & 30420   & 30023   & 30423  & 30420  & 34736 & 28588\\
                                  \hline
         \multirow{2}{*}{\makecell{Collision \\ (Advanced)}} & max   & 0.9999 & 100592  & 100143  & 100275  & 100174   & 100196  & 100195  & 168345 & 99456\\  
                            & min    & 0.9999 & 20497  & 20130  & 20412   & 20312   & 20384   & 20383  & 23864 & 19788\\ 
                                  \hline  
          Anesthesia & n/a  & 0.9999 & 13131  & 11462  & 11683   & 11658   & 11724   & 11722  & 13948 & 11288\\
			
             \specialrule{.12em}{.05em}{.05em}

		\end{tabular}
	 \end{center}
    \end{adjustwidth}
      \caption{Samples size comparison for confidence interval computation obtained via ProbReach, with solver  \textbf{$\delta$} precision equal to $10^{-3}$ and interval size equal to $10^{-2}$,  \textbf{Type} -  extremum type and  \textbf{\textit{c}} - confidence level.} \label{table:ResSam2}
	\end{table}

        \begin{figure}[!htb]
    \begin{adjustwidth}{-.6in}{-.6in}  
   \begin{minipage}{0.58\textwidth}
     \centering
     \includegraphics[width=1\linewidth]{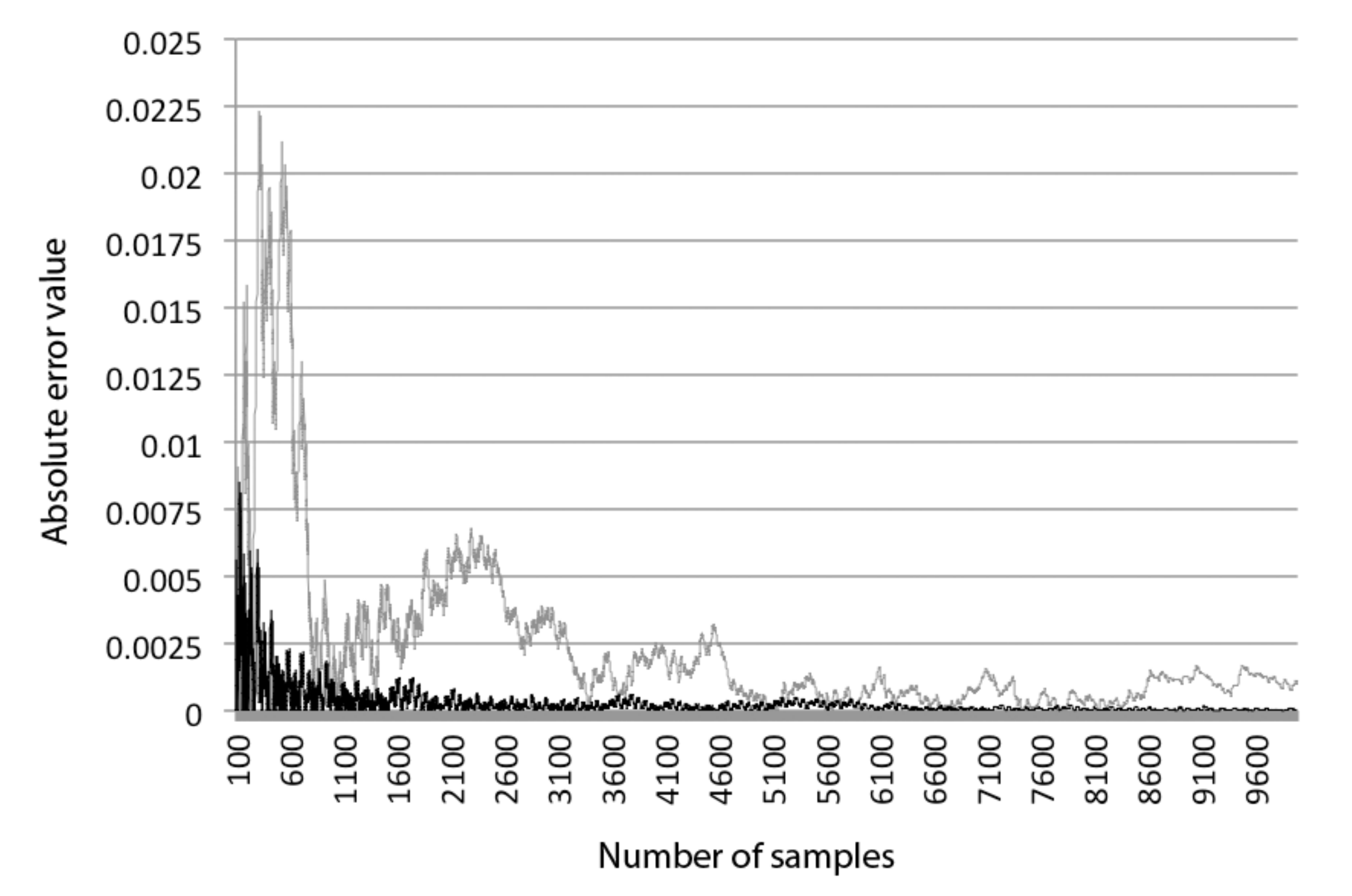}
     \caption{ MC (grey line) and QMC (black line) absolute error with respect to the number of samples. Model: Good, type - max.} \label{fig:GoodMax}
     \label{Fig:Data1}
   \end{minipage}\hfill
   \begin{minipage}{0.58\textwidth}
     \centering
     \includegraphics[width=1\linewidth]{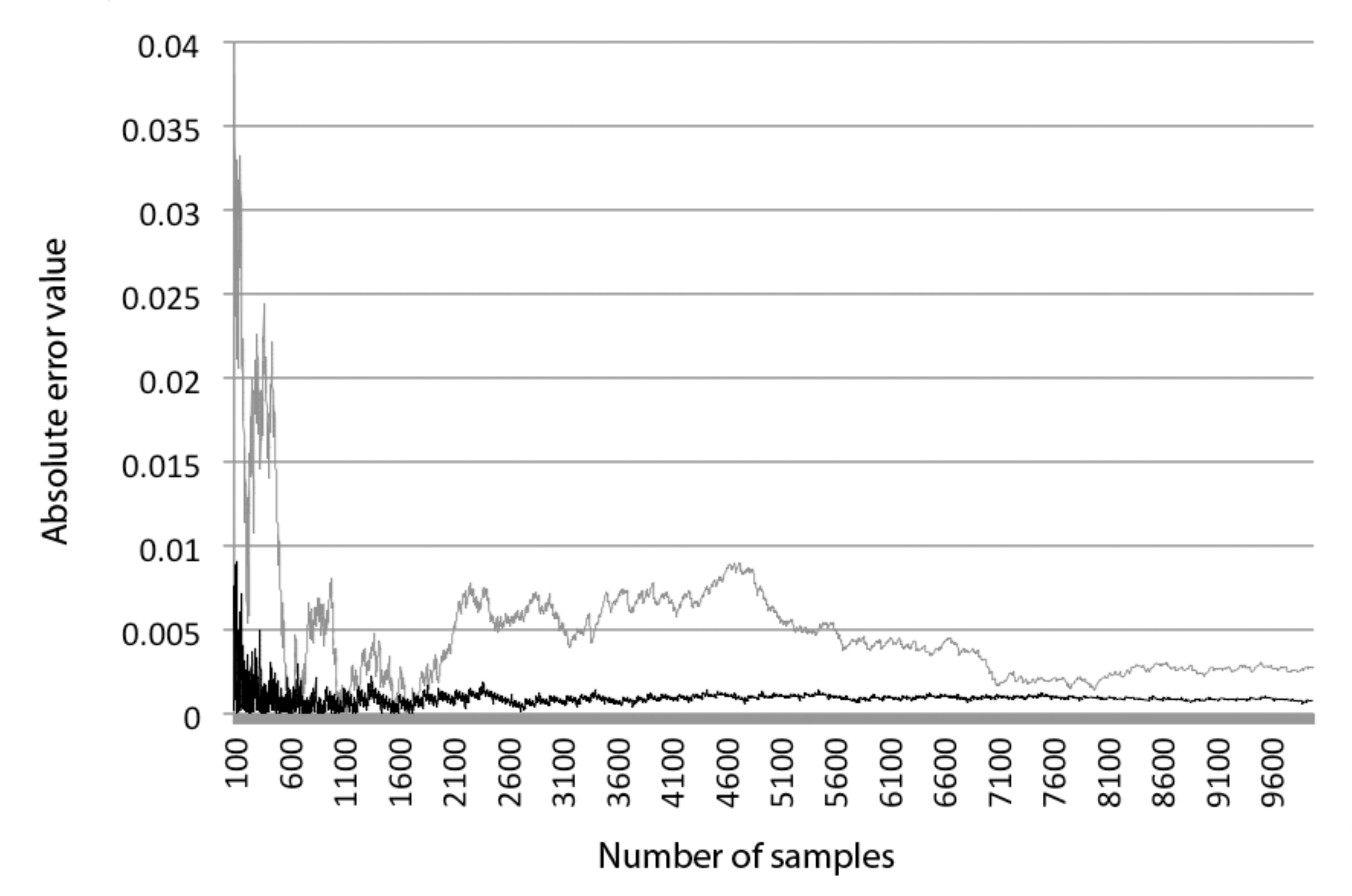}
     \caption{ MC (grey line) and QMC (black line) absolute error with respect to the number of samples, Model: Good, type - min.} \label{fig:GoodMin}
     \label{Fig:Data2}
   \end{minipage}
    \end{adjustwidth}
\end{figure}

  \begin{figure}[!htb]
    \begin{adjustwidth}{-.6in}{-.6in}  
   \begin{minipage}{0.58\textwidth}
     \centering
     \includegraphics[width=1\linewidth]{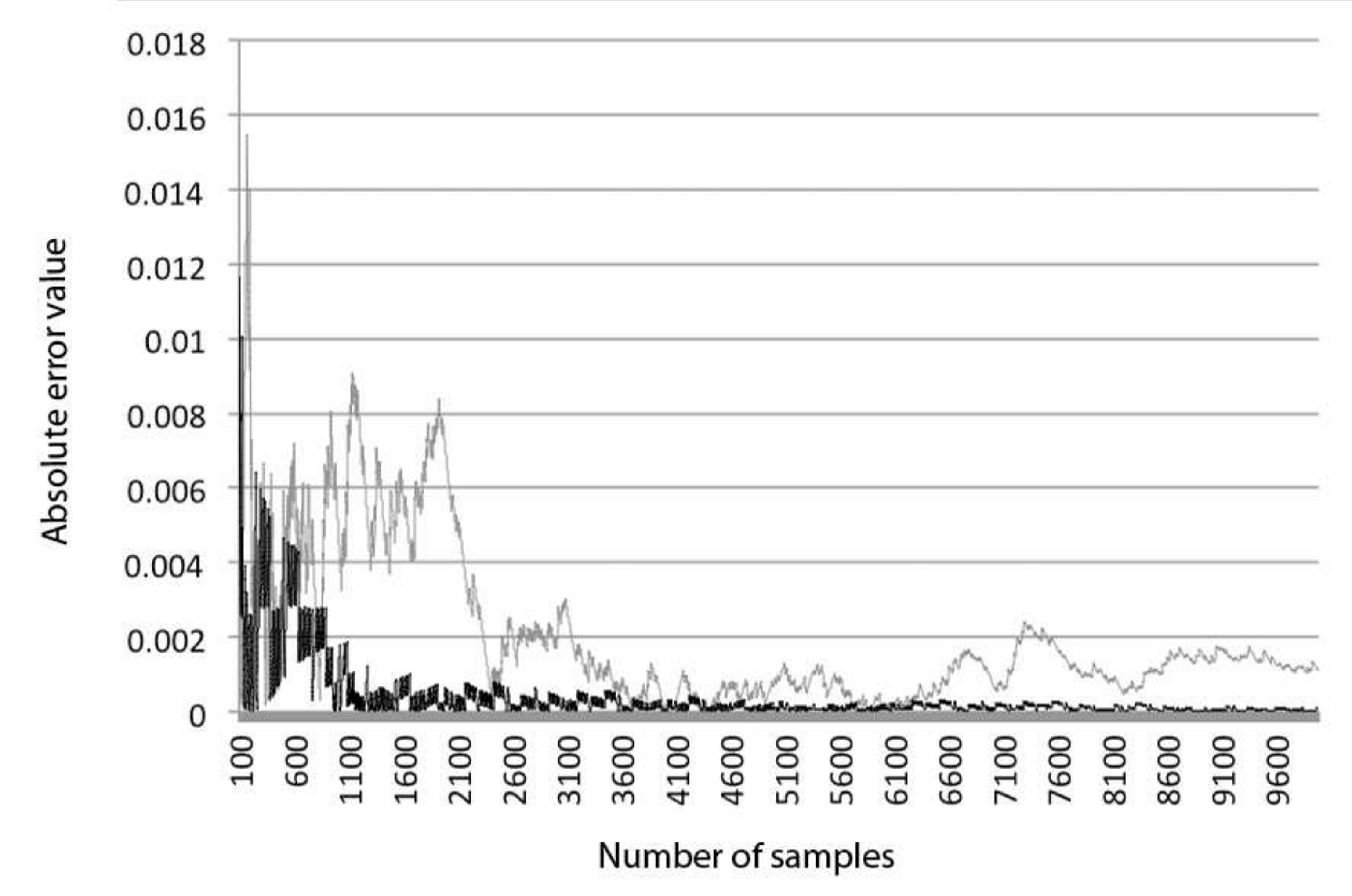}
     \caption{MC (grey line) and QMC (black line) absolute error with respect to the number of samples. Model: Bad, type - max.} \label{fig:BadMax}
     \label{Fig:Data1}
   \end{minipage}\hfill
   \begin{minipage}{0.58\textwidth}
     \centering
     \includegraphics[width=1\linewidth]{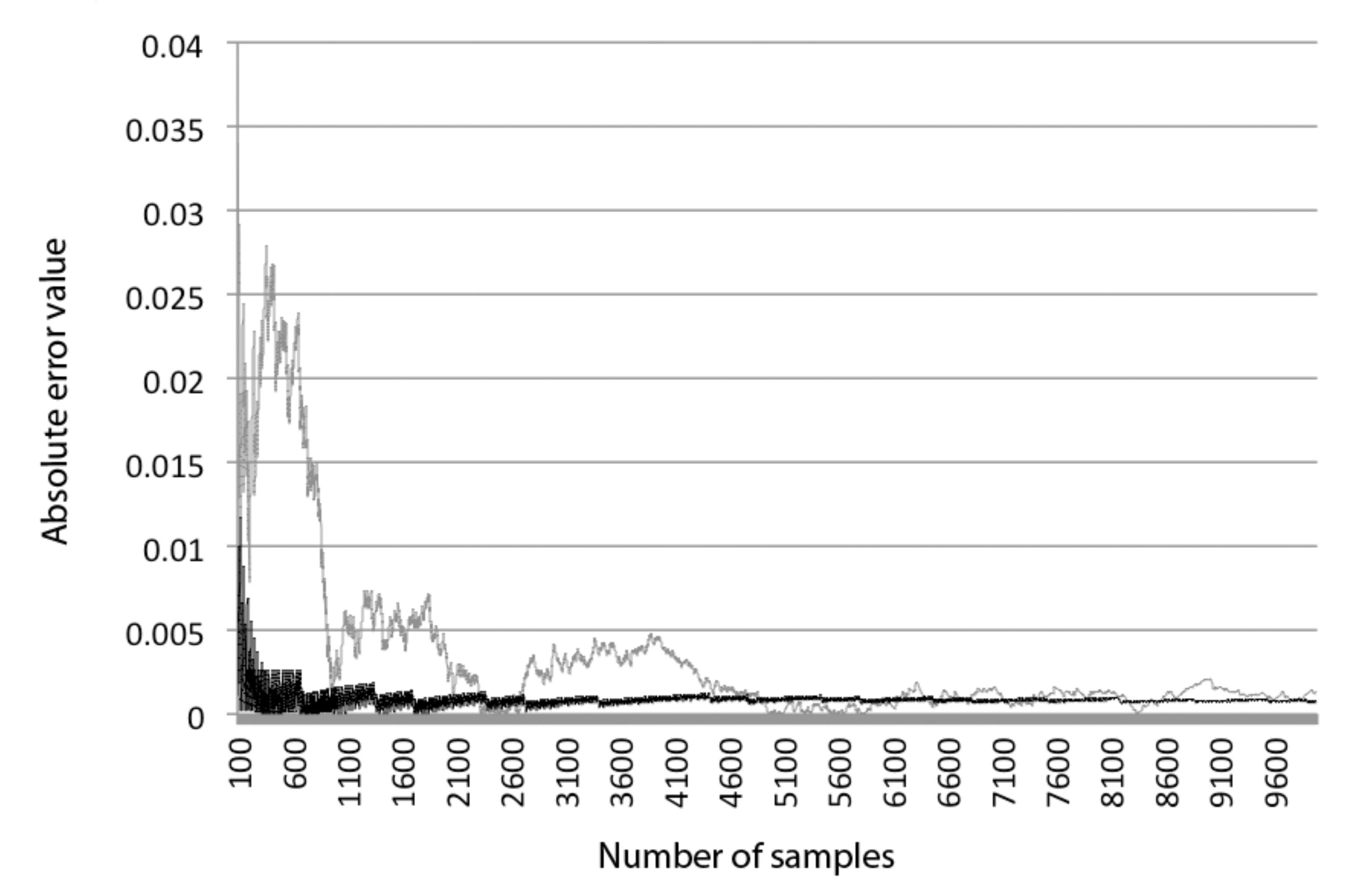}
     \caption{MC (grey line) and QMC (black line) absolute error with respect to the number of samples. Model: Deceleration, type - max.} \label{fig:DecMax}
     \label{Fig:Data2}
   \end{minipage}
    \end{adjustwidth}
\end{figure}

      \begin{figure}[!htb]
    \begin{adjustwidth}{-.6in}{-.6in}  
   \begin{minipage}{0.58\textwidth}
     \centering
     \includegraphics[width=1\linewidth]{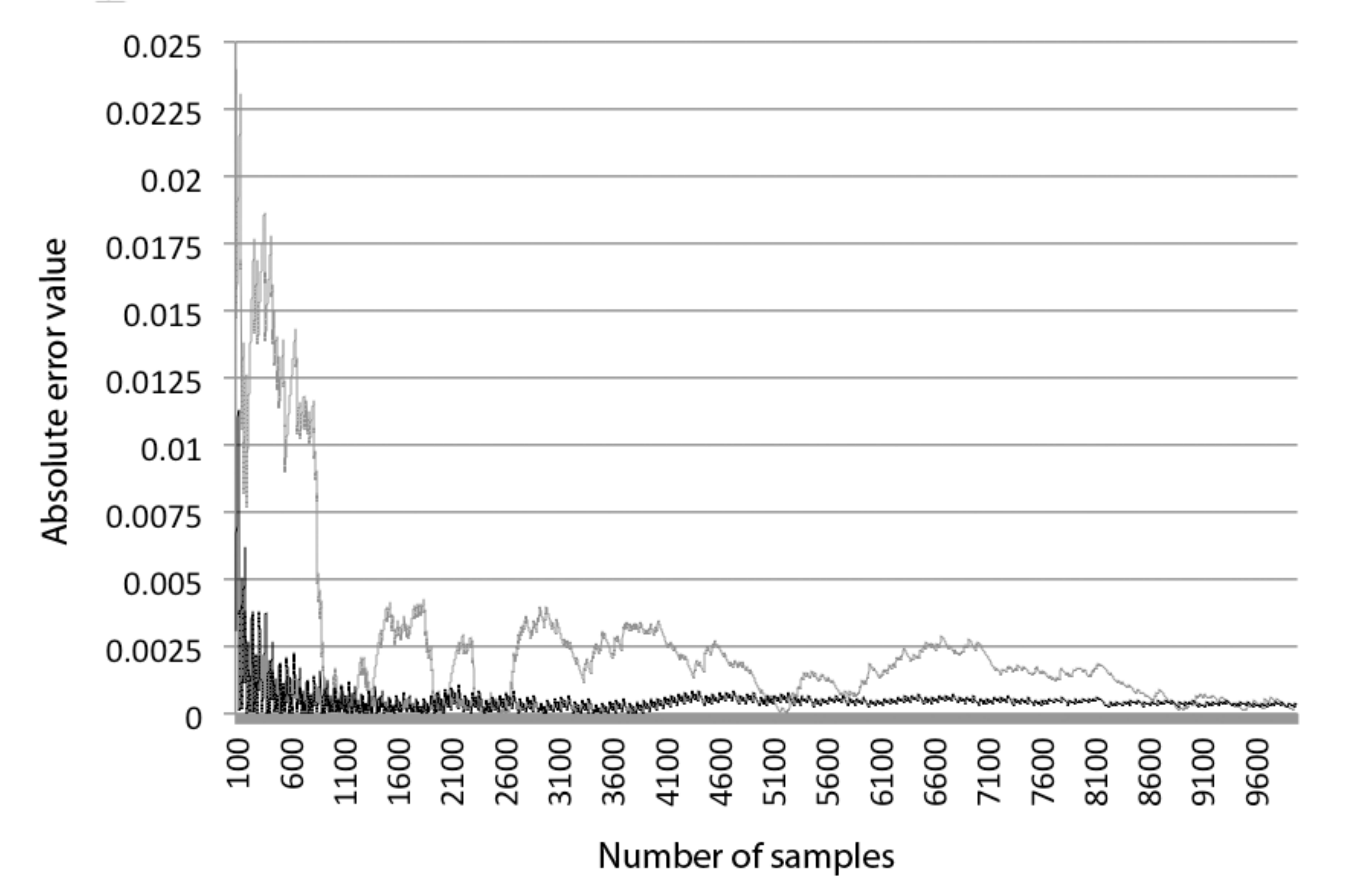}
     \caption{MC (grey line) and QMC (black line) absolute error with respect to the number of samples. Model: Deceleration, type - min. }
     \label{Fig:Data1}
   \end{minipage}\hfill
   \begin{minipage}{0.58\textwidth}
     \centering
     \includegraphics[width=1\linewidth]{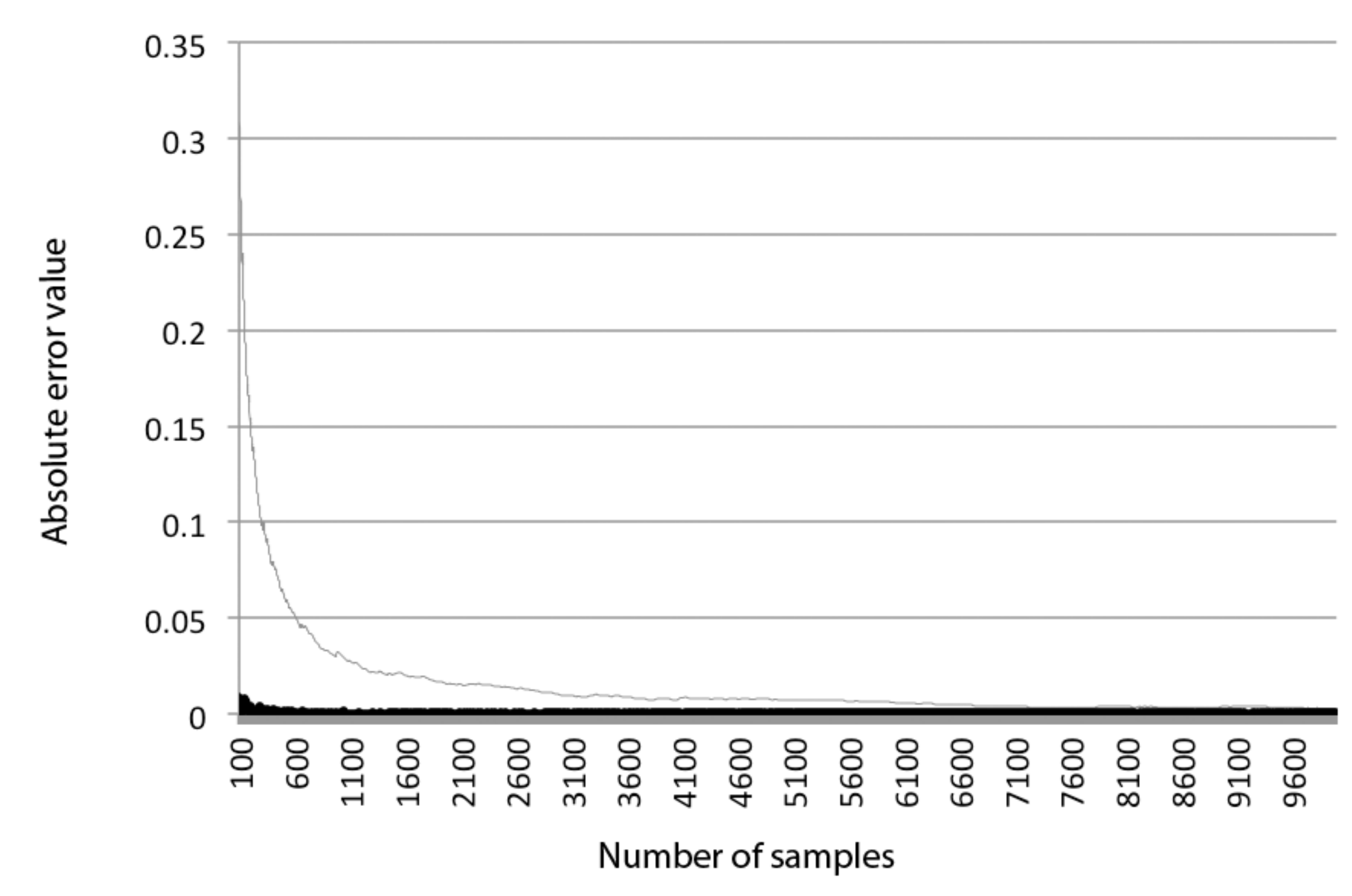}
     \caption{MC (grey line) and QMC (black line) absolute error with respect to the number of samples. Model: Collision basic, type - max.}
     \label{Fig:Data2}
   \end{minipage}
    \end{adjustwidth}
\end{figure}

 \begin{figure}[!htb]
    \begin{adjustwidth}{-.6in}{-.6in}  
   \begin{minipage}{0.58\textwidth}
     \centering
     \includegraphics[width=1\linewidth]{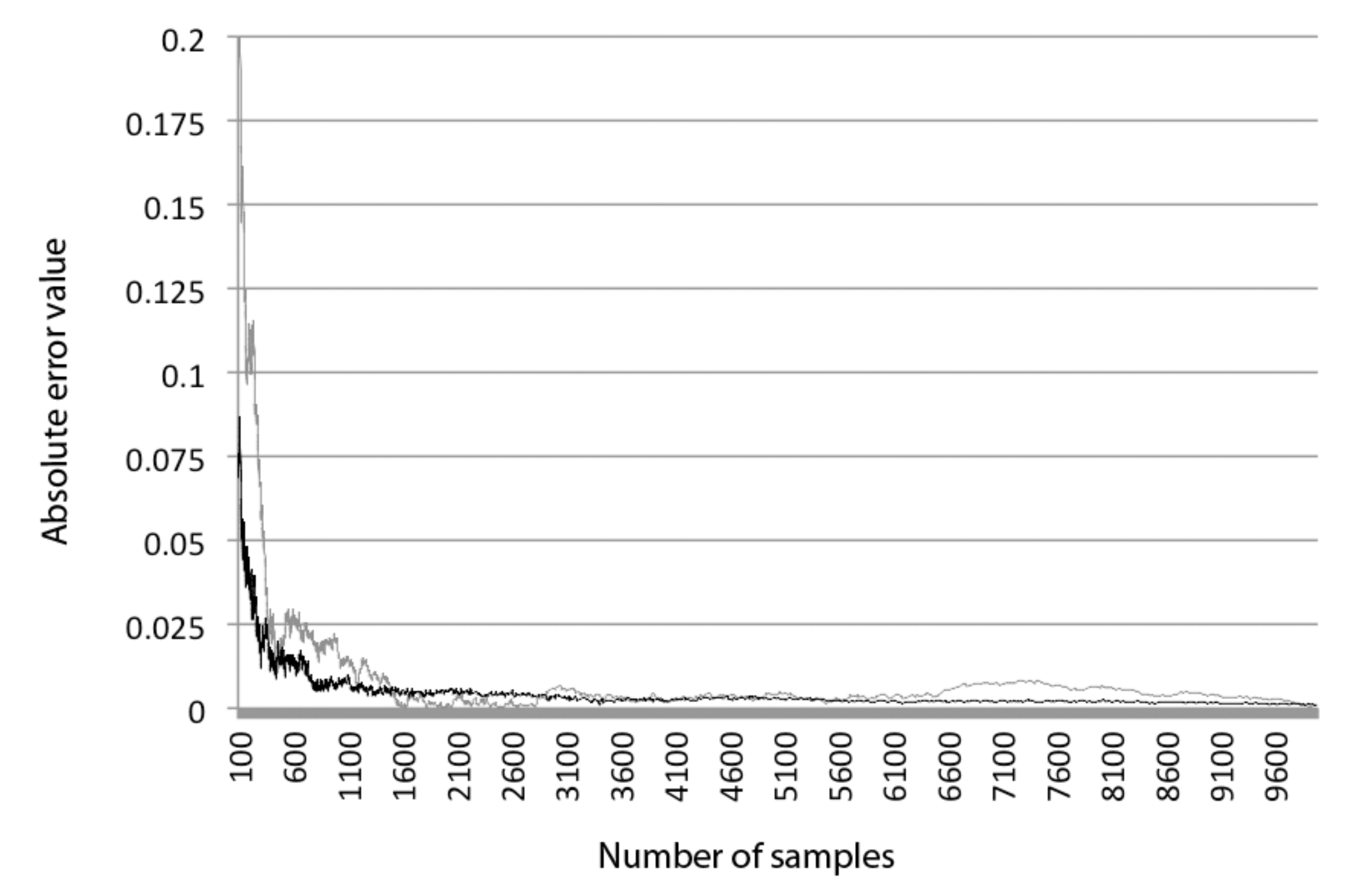}
     \caption{MC (grey line) and QMC (black line) absolute error with respect to the number of samples. Model: Collision extended, type - max.} \label{fig:ColExtMax}
     \label{Fig:Data1}
   \end{minipage}\hfill
   \begin{minipage}{0.58\textwidth}
     \centering
     \includegraphics[width=1\linewidth]{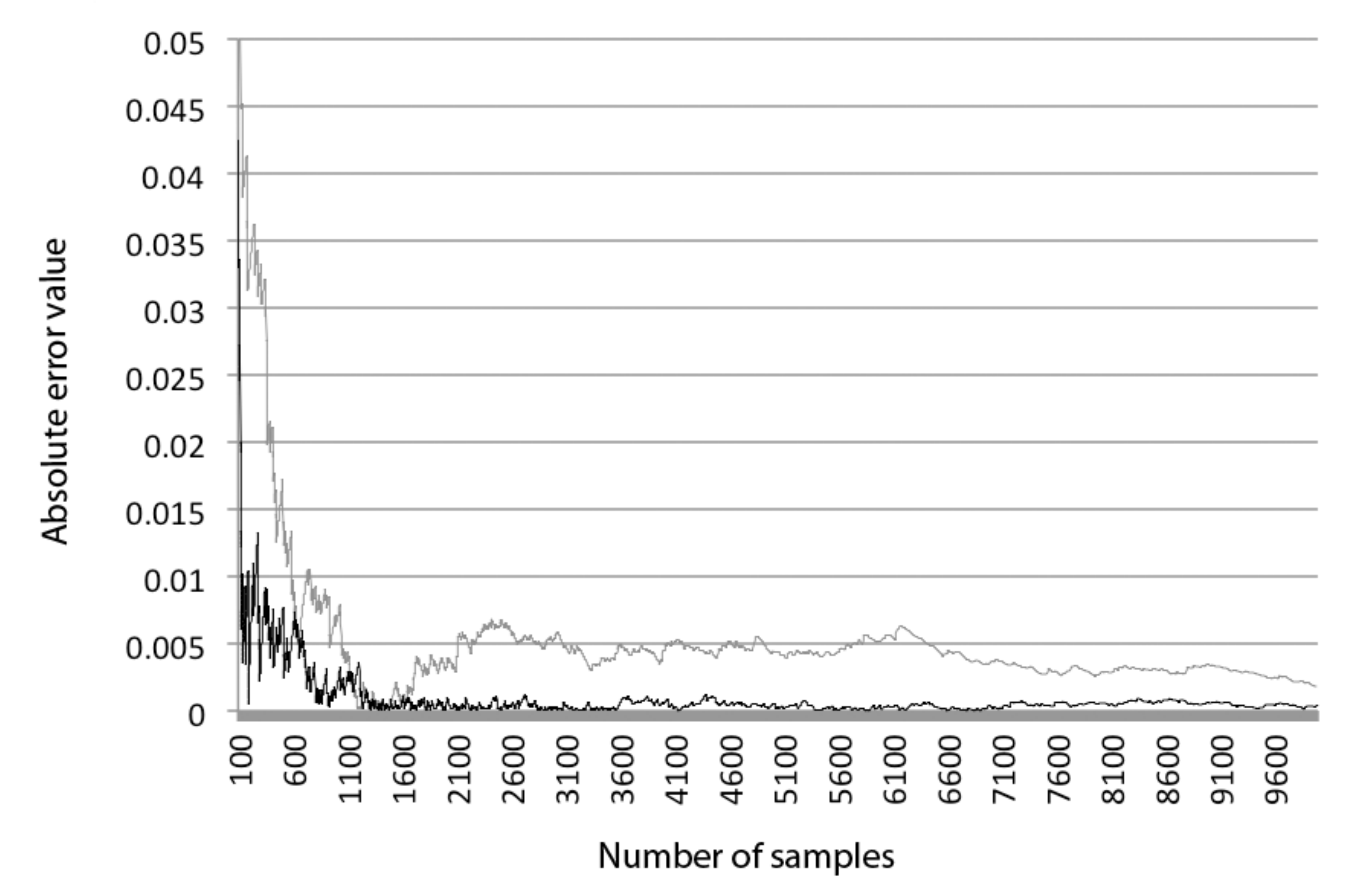}
     \caption{MC (grey line) and QMC (black line) absolute error with respect to the number of samples. Model: Collision extended, type - min.} \label{fig:ColExtMin}
     \label{Fig:Data2}
   \end{minipage}
    \end{adjustwidth}
\end{figure}

 \begin{figure}[!htb]
    \begin{adjustwidth}{-.6in}{-.6in}  
   \begin{minipage}{0.58\textwidth}
     \centering
       \includegraphics[width=1\linewidth]{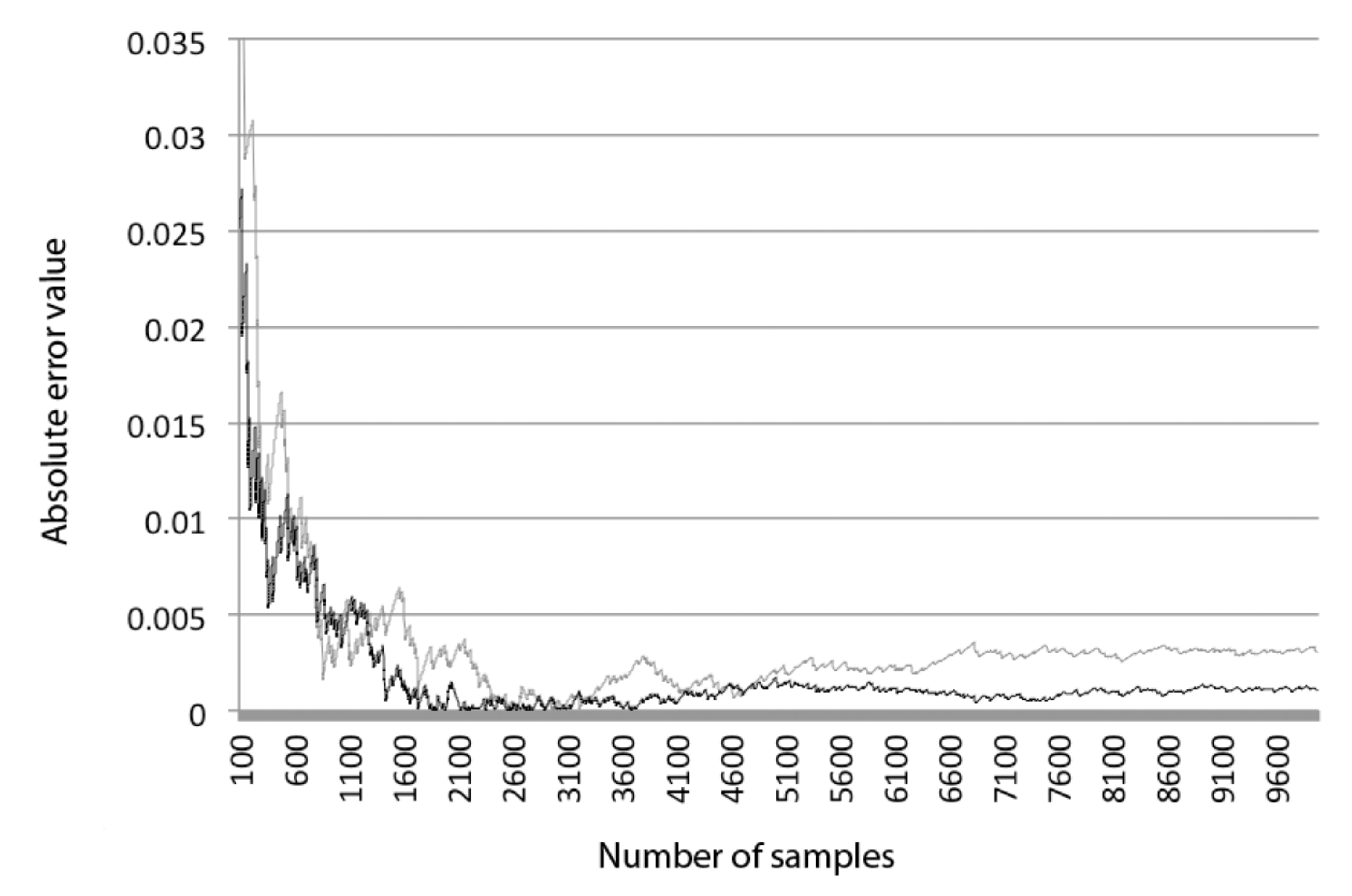}
     \caption{MC (grey line) and QMC (black line) absolute error with respect to the number of samples. Model: Collision advanced, type - min.} \label{fig:ColAdvMin}
     \label{Fig:Data1}
   \end{minipage}\hfill
   \begin{minipage}{0.58\textwidth}
     \centering
     \includegraphics[width=1\linewidth]{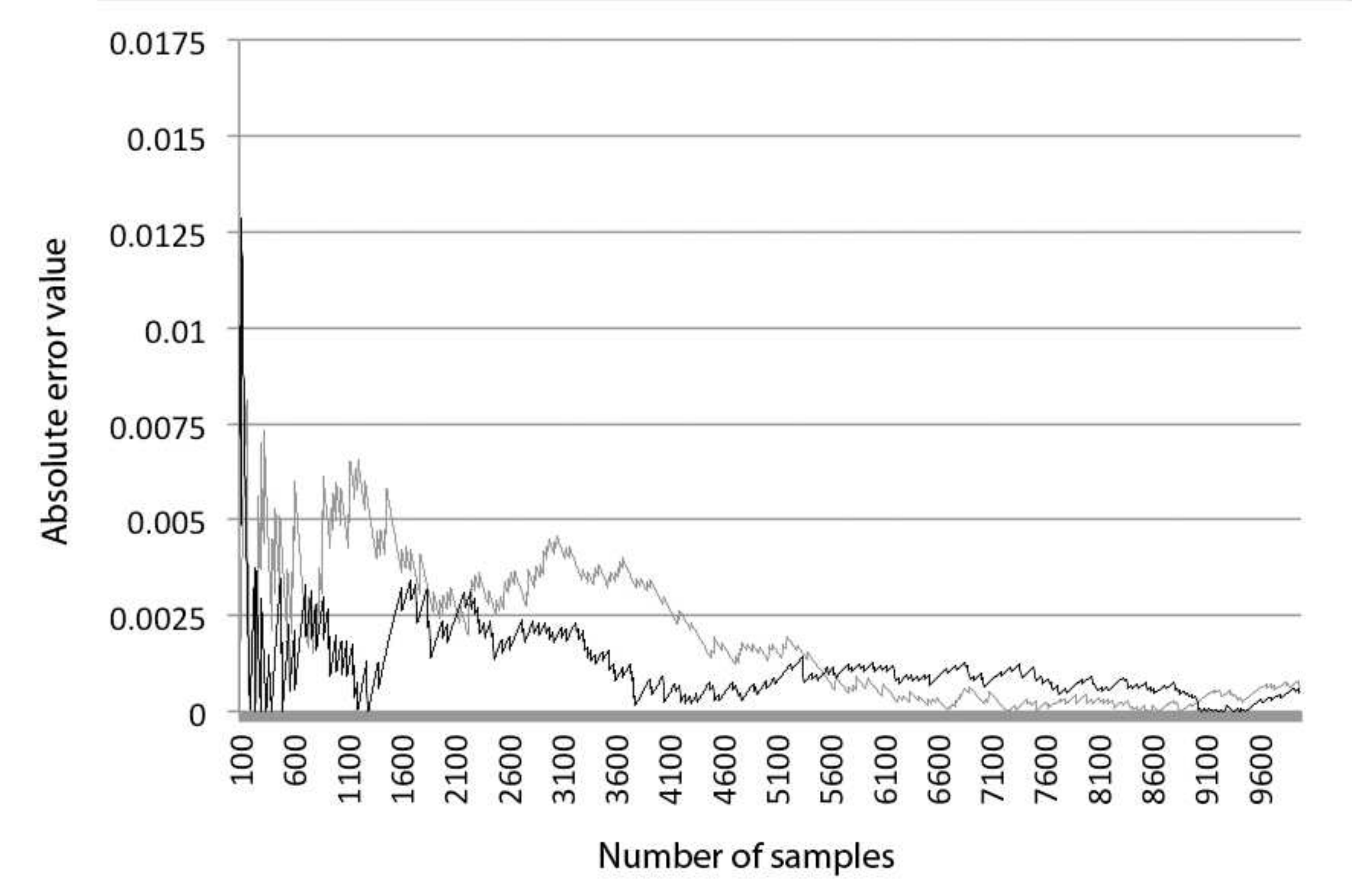}
 \newcommand{\tab}[1]{\hspace{.9\textwidth}}
     \caption{MC (grey line) and QMC (black line) absolute error with respect to the number of samples. Model: \tab\tab Anaesthesia. } \label{fig:Anest}
     \label{Fig:Data2}
   \end{minipage}
    \end{adjustwidth}
\end{figure}

\end{document}